\UseRawInputEncoding
\documentclass[sn-mathphys]{sn-jnl}% Math and Physical Sciences Reference Style
\usepackage{amsmath,amssymb}
\usepackage{subfigure}
\usepackage[section]{placeins}
\usepackage{epstopdf}

\jyear{2021}%
\theoremstyle{thmstyleone}%
\theoremstyle{thmstyletwo}%
\theoremstyle{thmstylethree}%
\begin{document}

\title[Article Title]{Echoes of charged black-bounce spacetimes}

\author[1]{\fnm{S. R.} \sur{Wu}}\email{gs.srwu20@gzu.edu.cn}

\author[2] {\fnm{B. Q.} \sur{Wang}}\email{wangbingqian13@yeah.net}

\author[1] {\fnm{Dong} \sur{Liu}}

\author*[1]{\fnm{Z. W.} \sur{Long}} \email{zwlong@gzu.edu.cn}

\affil[1]{\orgdiv{College of Physics}, \orgname{Guizhou University}, \orgaddress{\city{Guiyang}, \postcode{550025}, \country{China}}}

\affil[2]{\orgdiv{College Pharmacy}, \orgname{Guizhou University of Traditional Chinese Medicine}, \orgaddress{\city{Guiyang}, \postcode{550025}, \country{China}}}

\abstract{In present work, the evolution of scalar field and electromagnetic field under the background of the charged black-bounce spacetimes are investigated, and we obtain an obvious echoes signal which appropriately reports the properties of the charged black-bounce spacetimes and disclose the physical reasons behind such phenomena. Furthermore, by studying the quasinormal ringdown, we analyze the three states of the charged black-bounce spacetimes in detail, our results show that the echoes signal only appears when $(\rvert{Q}\rvert \le m)$ and $(\rvert {l}\rvert > m+ \sqrt{m ^{2}-Q^{2} })$ in this spacetime, while when the parameters demand $(\rvert {Q}\rvert>m)$,
the echoes signal will be transformed into a quasinormal ringdown of the two-way traversable wormhole, and the charged black-bounce is a regular black hole with normal horizons by requiring $(\rvert {Q}\rvert \le m)$ and $(\rvert {l}\rvert < m-\sqrt{m ^{2}-Q^{2} })$.}




\maketitle

\section{Introduction}\label{sec1}

In recent years, the first measurements of the shadow image captured by Event Horizon Telescope (EHT) \cite{ret1,ret2,ret3,ret4,ret5,ret6} and the gravitational wave (GW) searches by the LIGO Scientific and Virgo collaboration for the merger of compact binaries\cite{ret7,ret8,ret9,ret10}energized physicists in black hole physics, as the most extremely compact objects characterized by the existence of event horizons, black holes are some of the most important models in GW observations. In particular, GW echoes from black holes can help us to study the properties of the compact object itself, and moreover, probably furnish an avenue to the nature of quantum gravity. The notion of GW echoes was proposed in the context of the GW ringdown signals, aiming to study the phenomena in the vicinity of the black holes and their exotic compact alternatives.

On the other hand, as the characteristic sound of black holes, the quasinormal models (QNMs) are dominantly governed by the spacetimes in the vicinity of the horizon. It is well known that once a black hole is perturbed \cite{ret11}, there are three stages that are responsible for the perturbations evolution in time. The first one is a relatively short period of initial outburst of radiation, the second one is a usually long period of damping proper oscillations which dominated by the QNM, and the third one is the power-law tail. Here, we focus on the second stage of the evolution of perturbations represented by QNM. By analyzing black hole perturbation equations, one can derive the QNMs, and the potential solution is expressed by  the pure ingoing wave at the event horizon and the pure outgoing wave at infinity \cite{ret12,ret13}. Moreover, by studying the black hole QNMs, we can disclose the frequencies of QNMs \cite{ret14,ret15}, which are not only related to the basic physical properties of the black hole but also may provide reliable evidence of the existence of the black holes \cite{ret16}. The QNMs of wormholes have drawn much attention. Poulami Dutta Roy et al \cite{ret17} studied a family of ultra-static Lorentzian wormholes and obtained the scalar QNMs, their results may be used as a template for further studies on the gravitational wave physics of exotic compact objects; in Ref.\cite{ret18}, they studied the phantom wormholes and observe obvious signals of echoes, their results showed that the dark energy equation of state has a clear imprint in echoes in wave perturbations; M.S.Churilova et al\cite{ret19} studied the QNM of regular black-hole/wormhole transition and obtained unique wormhole echoes picture near the threshold, their results showed that different parameters corresponding to different spacetime, and only the traversable wormhole spacetime background has a clear echo picture. In addition, more related works can been seen in Refs.\cite{ret20,ret21,ret22,ret23,ret24,ret25,ret26,ret27}.

Given the demonstrated existence of the black-bounce-Kerr geometries \cite{ret28} and the black-bounce-Schwarzschild \cite{ret29}, it is reasonably suspect that analogous black bounce variants of the Reissner-Nordstr{\"o}m spacetimes will exist. Physically, we can find some answers in Ref.\cite{ret30}, in which the regularizing procedure has been proposed, i.e., the radial coordinate $r$ in the Reissner-Nordstr{\"o}m metric is replaced by $\sqrt{r^{2}+l^{2}}$, here $l$ is the bounce parameter of the charged black-bounce associated with the Plank length. This geometry implies a regularizing procedure: the introduction of bounce parameter or a length scale is applied to the Reissner Nordstr{\"o}m geometry black hole, and which is  globally free from curvature singularities, passes all weak field observational tests and smoothly interpolates between regular black holes and charged traversable wormholes. Alternatively, we can comprehend the charged black-bounce spacetimes as: an electromagnetic charge is introduced to the black-bounce spacetimes \cite{ret31, ret32}, where the black-bounce can be derived from the Schwarzschild black hole in terms of the regularizing procedure. In this paper, we attempt to provide an investigation to the QNM of the charged black-bounce spacetimes and disclose the physical reasons behind such phenomena. Physically, we analyse the properties of spacetimes in the context of the perturbation of electromagnetic fields and scalar fields, and our results show that echoes signal appears on the condition that the parameters are constrained.

The paper is organized as follows. In Sec.\ref{sec2}, we introduce the charged black-bounce spacetimes, the equations of motion under the influence of scalar field and electromagnetic field  perturbation, meanwhile the corresponding effective potential is derived and the figure of the effective potential under the perturbations of scalar field and electromagnetic field are discussed. In Sec.\ref{sec3}, the time domain integration method is introduced. In Sec.\ref{sec4}, the time-domain profiles of the scalar field and electromagnetic field in the context of the black hole and the wormhole by configurating different spacetime parameters are analyzed, including the different ringdown behaviors. In Sec.\ref{sec5}, the QNM frequencies of the charged black-bounce spacetimes are reported. Finally, the work is summarized in Sec.\ref{sec6}.

\section{Scalar field  and electromagnetic field  perturbation in the charged black-bounce spacetimes}\label{sec2}
We introduce the charged black-bounce spacetimes, its metric is given by \cite{ret33}

\begin{equation} d s^{2}= -A(r)d t^{2}+A^{-1}(r)d r^{2}+(r^{2}+l^{2})d{\Omega}^{2}, \label{eq1}   \end{equation}

\begin{equation}  A(r)= 1- \frac{2m}{\sqrt{r^{2}+l^{2}}} + \frac{Q^{2}}{r^{2}+l^{2}}, \label{eq2}   \end{equation}
here $m$ is the mass and $Q$ is the charge of the charged black-bounce respectively. $r\in (-\infty,+\infty)$ and $d{\Omega}^{2}$ represents the line element of a unit 2-sphere. We can see that natural domains for the angular and temporal coordinates are unaffected by the regularisation procedure, then compared with the Reissner-Nordstr{\"o}m spacetimes, the natural domain of the $r$ coordinate expands from $r\in (0,+\infty)$ to $r\in (-\infty,+\infty)$. It should be noted that different parameters correspond to different spacetimes, (a) the traversable wormhole:$(\rvert {Q}\rvert \le m)$ and $(\rvert {l}\rvert > m\pm \sqrt{m ^{2}-Q^{2} })$; (b) the two-way  traversable wormhole:$(\rvert {Q}\rvert>m)$; (c) the regular black hole:$(\rvert {Q}\rvert \le m)$ and $(\rvert {l}\rvert < m- \sqrt{m ^{2}-Q^{2} })$. Furthermore, the horizons are located at
\begin{equation} r_{H}= S_1\sqrt{(m+S_2 \sqrt{m ^{2}-Q^{2}} )^{2}-l ^{2} }, \label{eq3}   \end{equation}
where $S_1=S_2=\pm 1$, in particularly, $S_1=1$ corresponds to our universe, $S_1=-1$ labels the copy of our universe, $S_2=1$  indicates outer horizon and $S_2=-1$ represents inner horizon, respectively.

The general covariant K-G equation of scalar field can be expressed as
\begin{equation} \frac{1}{\sqrt {-g}}\partial_{\mu}(\sqrt {-g}g^{\mu\nu}\partial_{\nu}\Psi )=0 , \label{eq4} \end{equation}
in order to reduce the above equation, we assume the scalar field separated in the standard form \cite{ret34}
\begin{equation} \Psi(t,r,{\theta},{\phi})=\frac{1}{\sqrt{r ^{2}+l^{2}}}\Phi_{0}(r,t)Y^{m}_{l_{0}}({\theta},{\phi}),\label{eq5} \end{equation}
where $Y^{m}_{l_{0}}$ is spherical harmonic function of degree $l_{0}$ related to the angular coordinates $\theta$,$\phi$, and it demands

\begin{equation}[\frac{1}{sin{\theta}}\frac{\partial}{\partial{\theta}}(sin{\theta}\frac{\partial}{\partial{\theta}})+\frac{1}{sin^{2}{\theta}}   \frac{\partial^{2}}{\partial{\phi}^{2}}]Y^{m}_{l_{0}}({\theta},{\phi})=-l_{0}(l_{0}+1)Y^{m}_{l_{0}}({\theta},{\phi}), \label{eq6}       \end{equation}
combination the separated scalar function Eq.\eqref{eq5} and Eq.\eqref{eq4} obtains the following Regge-Wheeler wave equation relating the radial dependence of the scalar function and time
\begin{equation}[\frac{\partial^{2}}{\partial{t}^{2}}-\frac{\partial^{2}}{\partial r^{2}_{*}}+V_{0}(r)]\Phi_{0}(r,t)=0, \label{eq7}       \end{equation}
 with $V_{0}= -r ^{2}A^{2}(r ^{2}+l^{2})^{-2}+Ar[2mr(r^{2}+l^{2})^{-\frac{3}{2}}-2rQ^{2}(r ^{2}+l^{2})^{-2}](r ^{2}+l^{2})^{-1}+A^{2}(r ^{2}+l^{2})^{-1}+\frac{Al_0 (l_0+1)}{r ^{2}+l^{2}} $.

Furthermore, in view of the formalism of the charged black-bounce spacetimes, here the perturbed metric will be assumed to be axially symmetric, thus we can express the general metric which is time-dependent and axially symmetric as
\begin{equation}
d s^{2}=e^{2 v}(d t)^{2}-e^{2 \psi}\left(d \phi-W d t-q_{2} d x^{2}-q_{3} d x^{3}\right)^{2}-e^{2 \mu_{2}}\left(d x^{2}\right)^{2}-e^{2 \mu_{3}}\left(d x^{3}\right)^{2}, \label{eq8}
\end{equation}
thus in Eq.\eqref{eq1}
\begin{equation}
\begin{aligned}
 & e^{2 v}=e^{-2 \mu_{2}}=A=\frac{\Delta}{R^{2}},  \Delta=r^{2}+l^{2}-2 m \sqrt{r^{2}+l^{2}}+Q^{2}, \\  &R=\sqrt{r^{2}+l^{2}}, W=q_{2}=q_{3}=0, \label{eq9}
\end{aligned}
\end{equation}
and we take the coordinates $x^{2}=r$ and $x^{3}=\theta$.

The nonzero component of electric field strength of unperturbed metric is taken in the form
\begin{equation}
F_{02}=-\frac{Q}{R^{2}}, \label{eq10} \end{equation}
according to the linearized Maxwell's equations, since we just consider the magnetic field perturbation, only four relevant equations are listed as

\begin{equation}
\begin{aligned}
&\left(e^{\psi+\mu_{2}} F_{12}\right)_{, 3}+\left(e^{\psi+\mu_{3}} F_{31}\right)_{, 2}=0 \\
&\left(e^{\psi+v} F_{01}\right)_{, 2}+\left(e^{\psi+\mu_{2}} F_{12}\right)_{, 0}=0 \\
&\left(e^{\psi+v} F_{01}\right)_{, 3}+\left(e^{\psi+\mu_{3}} F_{13}\right)_{, 0}=0 \\
&\left(e^{\mu_{2}+\mu_{3}} F_{01}\right)_{, 0}+\left(e^{v+\mu_{3}} F_{12}\right)_{, 2}+\left(e^{v+\mu_{2}} F_{13}\right)_{, 3}= \\
&e^{\psi+\mu_{3}} F_{02} O_{02}+e^{\psi+\mu_{2}} F_{03} O_{03}-e^{\psi+v} F_{23} O_{23},
 \label{eq11}      \end{aligned}      \end{equation}
where $O_{j 0}=\frac{\partial q_{j}}{\partial t}-\frac{\partial W}{\partial x^{j}}$ and $O_{j B}=\frac{\partial q_{j}}{\partial x^{B}}-\frac{\partial q_{B}}{\partial x^{j}}(j, B=2,3)$.

Here it should be noted that the first equation in Eq.\eqref{eq11} can be neglected since it just provides the integrability conditions for the next two equations, then we can rewrite this equation in spherical coordinates as
\begin{equation}
\begin{aligned}
&\left(R \sin \theta e^{v} F_{01}\right)_{, r}+R \sin \theta e^{-v} F_{12,0}=0 \\
&\left(R \sin \theta e^{v} F_{01}\right)_{, \theta}+R^{2} \sin \theta F_{13,0}=0 \\
&e^{-v} R F_{01,0}+\left(e^{v} R F_{12}\right)_{, r}+F_{13, \theta}=-Q \sin \theta\left(W_{, 2}-q_{2,0}\right).
\end{aligned} \label{eq12}    \end{equation}
The perturbed components of the Ricci tensor for charged black-bounce spacetimes are computed as follows
\begin{equation}
\begin{aligned}
&\delta R_{(a)(b)}=-2\left[\eta^{(n)(m)}\left(\delta F_{(a)(n)} F_{(b)(m)}+F_{(a)(n)} \delta F_{(b)(m)}\right)-\frac{\eta_{(a)(b)} Q \delta F_{02}}{R^{2}}\right] \\
&\delta R_{12}=2 \frac{Q}{R^{2}} \delta F_{01} \\
&\delta R_{02}=\delta R_{13}=0,
\end{aligned} \label{eq13}    \end{equation}
in view of the fact that only the axial perturbations are considered in this part, it is characterized by non-zero values $W$, $q_{2}$, $q_{3}$, $F_{01}$, $F_{12}$ and $F_{13}$, and the two equations governing axial perturbations comes from $R_{12}$ and $R_{13}$. Besides, it should be noted that before perturbations, $R_{12}=R_{13}=0$, after perturbations, neither of them is zero. Therefore we substitute the changes in the corresponding Ricci tensors and remain the functions $v$, $\psi$, $\mu_{2}$ and $\mu_{3}$ as same as before the perturbations, one can obtain
\begin{equation}
\begin{aligned}
&\left(R^{2} \sin ^{3} \theta e^{2 v} O_{23}\right)_{, 3}+R^{4} \sin ^{3} \theta O_{02,0}=4 R \sin ^{2} \theta e^{v} Q \delta F_{01} \\
&\left(R^{2} \sin ^{3} \theta e^{2 v} O_{23}\right)_{, 2}-R^{2} \sin ^{3} \theta e^{-2 v} O_{03,0}=0,  \label{eq14}
\end{aligned}    \end{equation}
eliminating $F_{12}$ and $F_{13}$ from Eq.\eqref{eq12} leads to
\begin{equation}
-e^{-v} R B_{, 0,0}+\left[e^{2 v}\left(\operatorname{Re}^{v} B\right)_{, r}\right]_{, r}+\frac{e^{v}}{R} \sin \theta\left(\frac{B_{, \theta}}{\sin \theta}\right)_{, \theta}=\sin ^{2} \theta Q\left(W_{, 2,0}-q_{2,0,0}\right),  \label{eq15}
\end{equation}
with $B=F_{01}sin\theta$.
By defining new function $\eta$
\begin{equation}
\eta(r, \theta, t)=R^{2} e^{2 v} \sin ^{3} \theta O_{23}=\Delta \sin ^{3} \theta\left(q_{2,3}-q_{3,2}\right),  \label{eq16}  \end{equation}
substituting Eq.\eqref{eq16} into Eq.\eqref{eq14}, and taking the time dependence of the perturbed values of $W$, $q_{2}$, $q_{3}$ and $O$ to be $e^{-i\omega t}$ where
$\omega$ is the frequency of the modes, we have
\begin{equation}
R^{4}\left(\frac{\Delta}{R^{4}} \frac{\partial \eta}{\partial r}\right)_{, r}+\sin ^{3} \theta\left(\frac{1}{\sin ^{3} \theta} \frac{\partial \eta}{\partial \theta}\right)_{, \theta}+\frac{\omega^{2} \eta}{\Delta} R^{4}=4 Q e^{v} R \sin ^{3} \theta\left(\frac{B}{\sin ^{2} \theta}\right)_{, \theta} ,  \label{eq17}
\end{equation}
combining Eq.\eqref{eq14} and Eq.\eqref{eq15} leads to
\begin{equation}
\left[e^{2 v}\left(\operatorname{Re}^{v} B\right)_{, r}\right]_{r}+\frac{e^{v}}{R} \sin \theta\left(\frac{B_{, \theta}}{\sin \theta}\right)_{, \theta}+\left(\omega^{2} e^{-v} R-\frac{4 Q^{2} e^{v}}{R^{3}}\right) B=-\frac{Q}{R^{4} \sin \theta} \frac{\partial \eta}{\partial \theta}.  \label{eq18}   \end{equation}
Next, we deal with the Eq.\eqref{eq17} and Eq.\eqref{eq18} by using the method of separation of variables
\begin{equation}
\begin{aligned}
&\eta(r, \theta)=\eta(r) C_{l_{0}+2}^{-\frac{3}{2}}(\theta) \\
&B(r, \theta)=\frac{B(r)}{\sin \theta} \frac{\partial C_{l_{0}+2}^{-\frac{3}{2}}(\theta)}{\partial \theta}=3 B(r) C_{l_{0}+1}^{-\frac{1}{2}}(\theta),
\end{aligned}
\label{eq19}   \end{equation}
with $C_{n}^{v}(\theta)$ are Gegenbauer polynomials, thus we derive the principal equation for axial perturbation as
\begin{equation}
\begin{aligned}
&\Delta\left(\frac{\Delta}{R^{4}} \frac{\partial \eta}{\partial r}\right)_{, r}-\frac{u^{2} \Delta}{R^{4}} \eta(r)+\omega^{2} \eta(r)=-4 Q e^{v} \frac{\Delta u^{2}}{R^{3}} B(r) \\
&{\left[e^{2 v}\left(\operatorname{Re}^{v} B\right)_{, r}\right]_{r}-\frac{e^{v}}{R}\left(u^{2}+2\right) B(r)+\left(\omega^{2} e^{-v} R-\frac{4 Q^{2} e^{v}}{R^{3}}\right) B=-\frac{Q}{R^{4}} \eta(r)}, \label{eq20}
\end{aligned}  \end{equation}
with $u^{2}=2n=(l_{0}-1)(l_{0}+2)$.

In order to facilitate the study, the variable $r$ will be replaced in terms of the tortoise coordinate defined by
\begin{equation}
\partial r=A \partial r_{*}=\frac{\Delta}{R^{2}} \partial r_{*},  \label{eq21} \end{equation}
$\eta(r)$ and $B(r)$ are redefined as
\begin{equation}
\begin{aligned}
&\eta(r)=R H_{2} \\
&B(r)=-\frac{H_{1}}{2 u} R^{-1} e^{-v}. \label{eq22}
\end{aligned}
\end{equation}
With these definitions, Eq.\eqref{eq20} becomes
\begin{equation}
\begin{aligned}
\Lambda^{2} H_{2}=\frac{\Delta}{R^{5}}\left(\begin{array}{l}
2 Q u H_{1}+\frac{-l^{4}+2 r^{2}\left(2 Q^{2}+r^{2}-3 m \sqrt{r^{2}+l^{2}}\right)+l^{2}\left(-Q^{2}+r^{2}+2 m \sqrt{r^{2}+l^{2}}\right)}{\left(r^{2}+l^{2}\right)^{\frac{3}{2}}}H_{2} \\
+u^{2} R H_{2}
\end{array}\right) \\
\Lambda^{2} H_{1}=\frac{\Delta}{R^{5}}\left(2 Q u H_{2}+\left(u^{2}+2\right) R H_{1}+\frac{4 Q^{2}}{R} H_{1}\right) ,  \label{eq23}
\end{aligned}
\end{equation}

here
\begin{equation} \Lambda^{2}=\frac{\partial^{2}}{\partial r_{*}^{2}}+\omega^{2},  \label{eq24}   \end{equation}
and $l=0$ the Eq.\eqref{eq23} approaches the Reissner-Nordstr\"om black hole expressions\cite{ret35}. Furthermore, the Eq.\eqref{eq23} can be considered as two one-dimensional wave equations
coupled by the interaction matrix
\begin{equation}
\Lambda^{2}\left(\begin{array}{l}H_{1} \\ H_{2}\end{array}\right)=\left(\begin{array}{ll}U_{11} & U_{12} \\ U_{21} & U_{22}\end{array}\right)\left(\begin{array}{l}H_{1} \\ H_{2}\end{array}\right),  \label{eq25}   \end{equation}
where
\begin{equation}
\begin{aligned}
&U_{11}=\frac{\Delta}{R^{5}}\left[\left(u^{2}+2\right) R+\frac{4 Q^{2}}{R}\right] \\
&U_{12}=U_{21}=\frac{\Delta}{R^{5}} 2 Q u \\
&U_{22}=\frac{\Delta}{R^{5}}[\frac{-l^{4}+\lambda+l^{2}(-Q^{2}+r^{2}+2 m \sqrt{r^{2}+l^{2}})}{(r^{2}+l^{2})^{\frac{3}{2}}}+u^{2} R], \label{eq26}
\end{aligned}
\end{equation}
with $\lambda=2 r^{2}(2 Q^{2}+r^{2}-3 m \sqrt{r^{2}+l^{2}})$.

Noted that the matrix can be diagonalized by a similarity transformation as
\begin{equation}
\Lambda^{2} Z_{i}=V_{i} Z_{i}(i=1,2), \label{eq27}  \end{equation}
where
\begin{equation}
V_{1}=\frac{1}{2}\left(U_{11}+U_{22}+\sqrt{\left(U_{11}-U_{22}\right)^{2}+4 U_{12}^{2}}\right),  \label{eq28}      \end{equation}

\begin{equation}
V_{2}=\frac{1}{2}\left(U_{11}+U_{22}-\sqrt{\left(U_{11}-U_{22}\right)^{2}+4 U_{12}^{2} }\right).    \label{eq29}    \end{equation}

As stated before, the symbol $r_{*}$ is the tortoise coordinate which is given by
\begin{equation}r_{*}=\int \frac{dr}{A(r)}, \label{eq30} \end{equation}
and the lapse function $A(r)$ decides the event horizon location, i.e., $r_{*}$ $\rightarrow$ $-\infty$ as $r$ $\rightarrow$ the event horizon of the black hole from infinity, thus we have $r_{*}$ $\in$  ($-\infty$,$+\infty$) for $r$ $\in$ ($r_{H}$,$+\infty$), therefore the Regge Wheeler wave-like equation (Eq.\eqref{eq7} and Eq.\eqref{eq27}) is restricted to the regions located outside the event horizon, $r>r_{H}$. Moreover, the asymptotic regions of the wormhole correspond to $r_{*}$ $\rightarrow$ $\pm\infty$ when the horizons vanish, we express $\pm r_{*}$ in order to cover the two asymptotic regions connected by the throat, this proceture needs a thin shell joining two copies of the same geometry \cite{ret36}. Meanwhile, it is also worth pointing that the coordinate $r_{*}$ is monotonically increasing with $r$, which appropriately avoids the horizon and more conveniently to discuss the QNMs.

\begin{figure}[htbp]
\begin{tabular}{cc}
\begin{minipage}[t]{0.45\linewidth}
\centerline{\includegraphics[width=6.0cm]{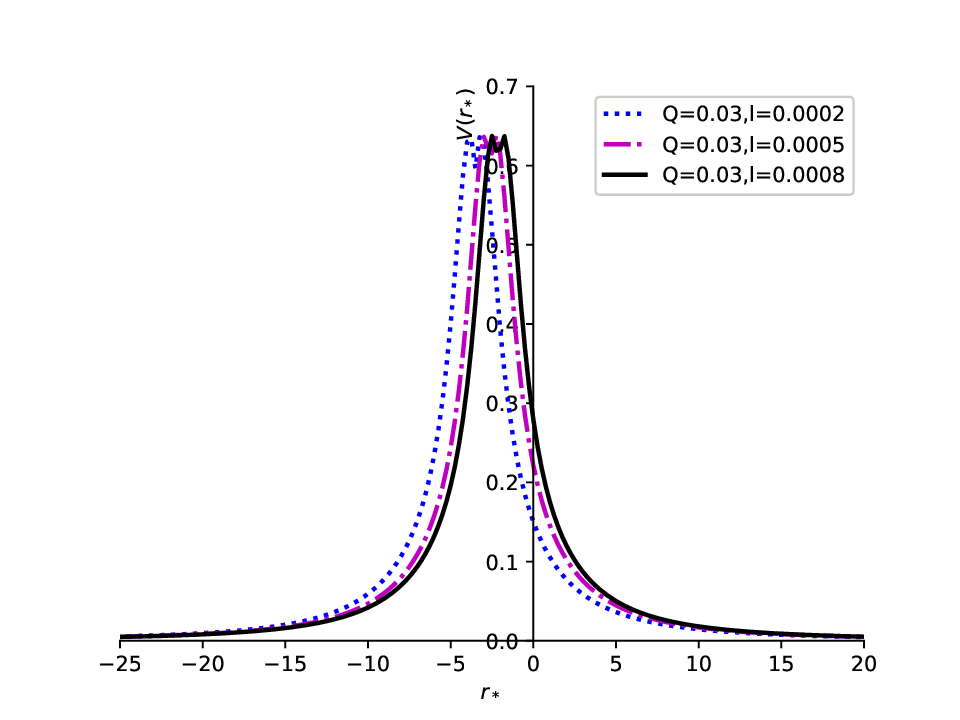}}
\centerline{(a)}
\end{minipage}
\hspace{7mm}
\begin{minipage}[t]{0.45\linewidth}
\centerline{\includegraphics[width=6.0cm]{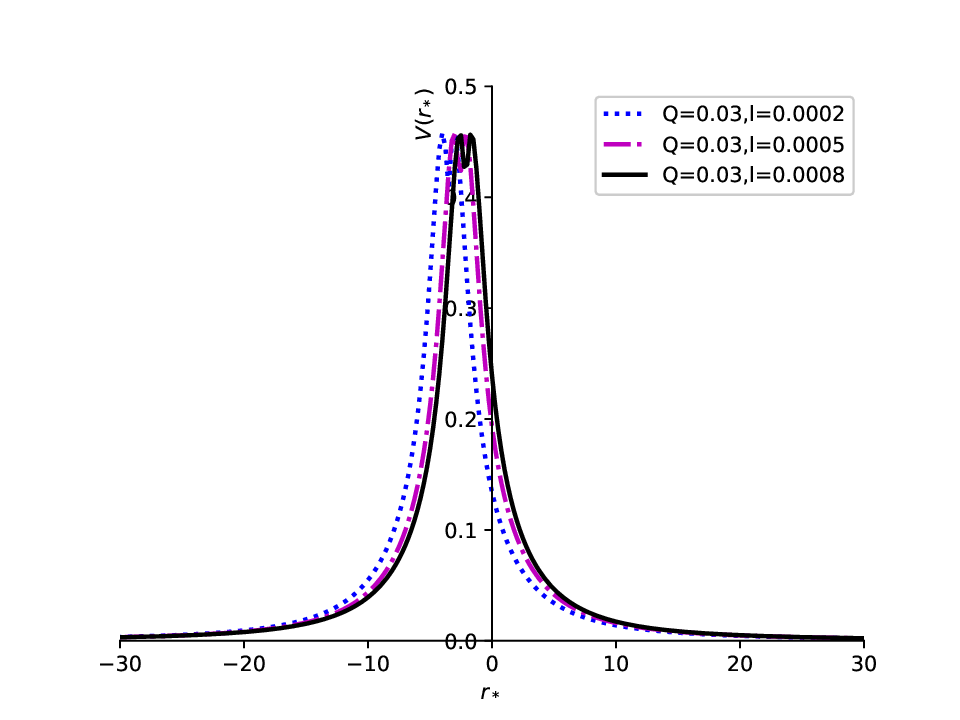}}
\centerline{(b)}
\end{minipage}
\end{tabular}
\caption{(a)The effective potentials as a function of tortoise coordinate $r_{*}$ for perturbations of the scalar field on the regular black hole spacetime with $m=0.5$, $l_{0}=1$;
(b)The effective potential($V_{1}$) as a function of tortoise coordinate $r_{*}$ for perturbations of the electromagnetic field on the regular black hole spacetime with $m=0.5$, $l_{0}=1$.}
\label{fig1}
\end{figure}

\begin{figure}[htbp]
\begin{tabular}{cc}
\begin{minipage}[t]{0.45\linewidth}
\centerline{\includegraphics[width=6.0cm]{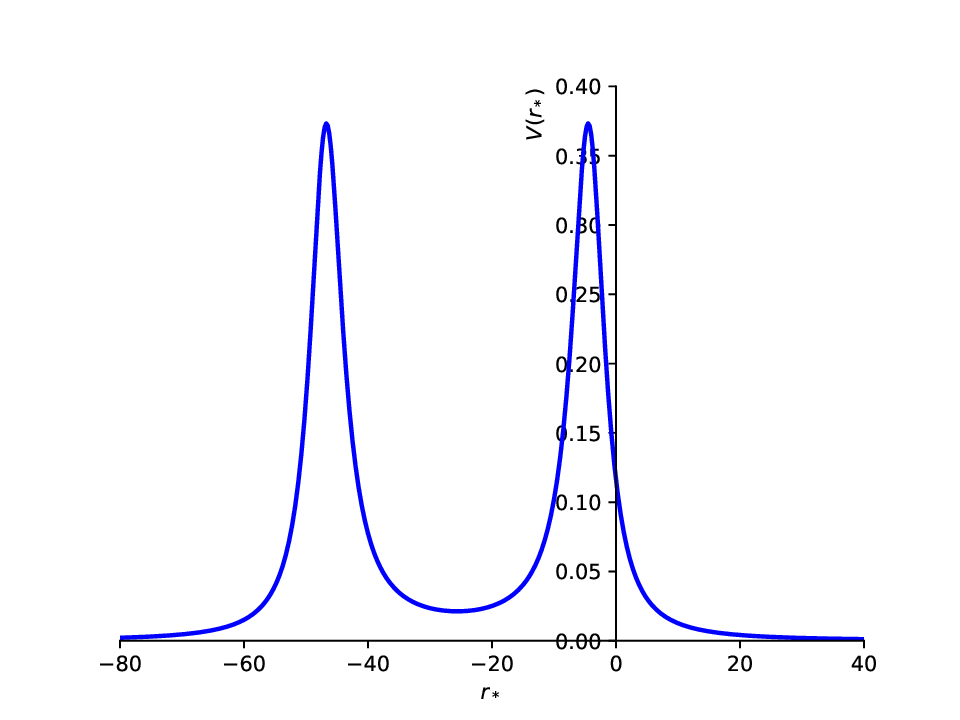}}
\centerline{(a)}
\end{minipage}
\hspace{7mm}
\begin{minipage}[t]{0.45\linewidth}
\centerline{\includegraphics[width=6.0cm]{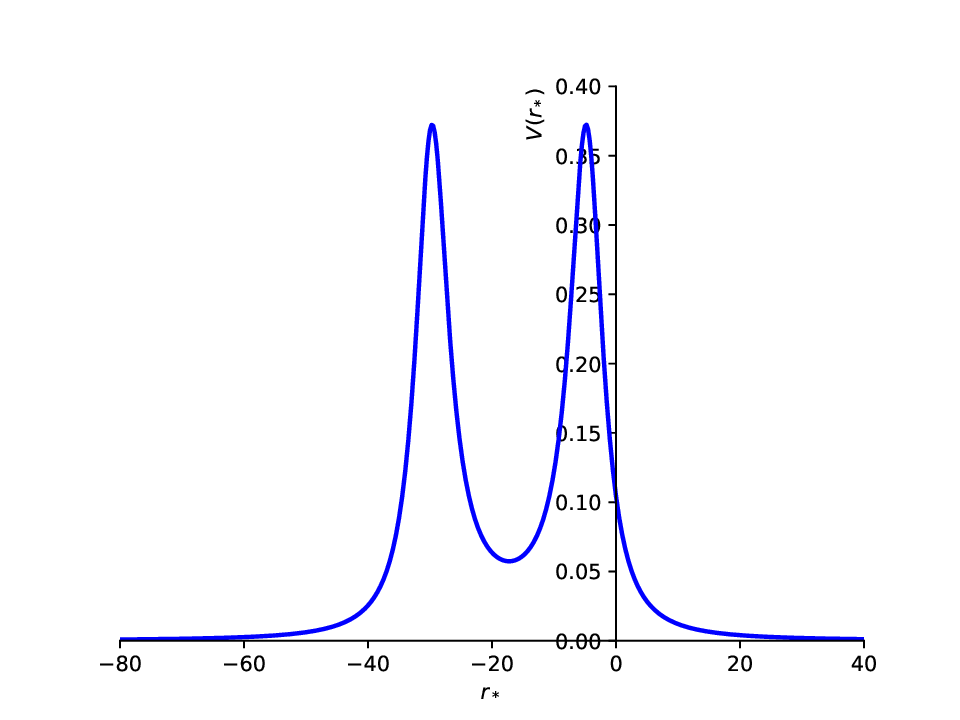}}
\centerline{(b)}
\end{minipage}
\hspace{7mm}
\\
\begin{minipage}[t]{0.45\linewidth}
\centerline{\includegraphics[width=6.0cm]{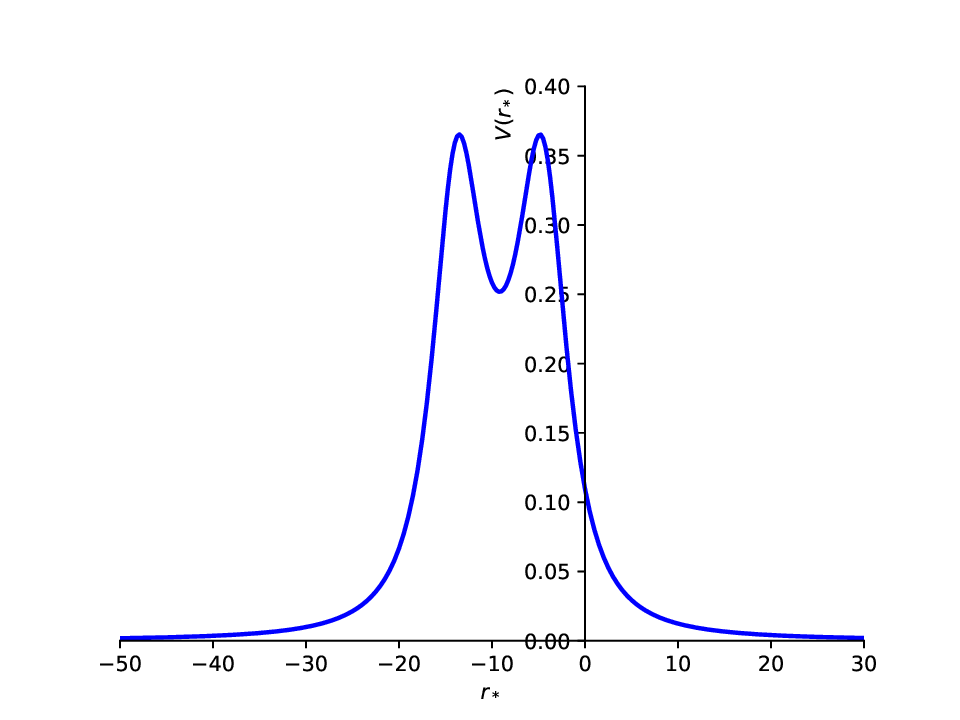}}
\centerline{(c)}
\end{minipage}

\hspace{7mm}
\begin{minipage}[t]{0.45\linewidth}
\centerline{\includegraphics[width=6.0cm]{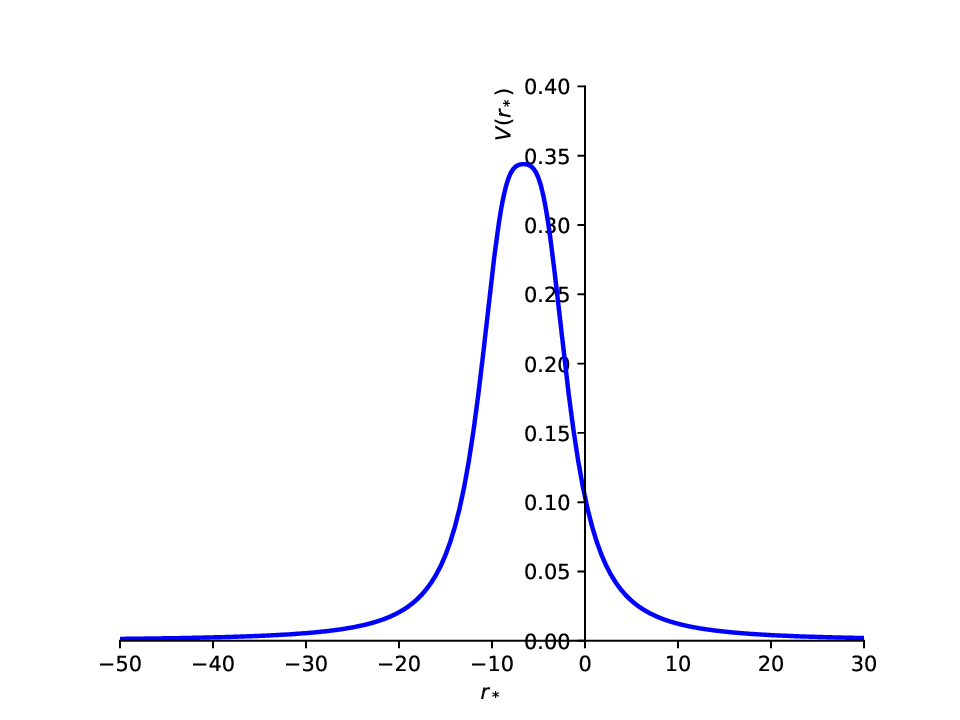}}
\centerline{(d)}
\end{minipage}
\end{tabular}
\caption{(a)The effective potentials as a function of tortoise coordinate $r_{*}$ for perturbations of the scalar field on the traversable wormhole spacetime with $l=1.01$; (b)The effective potentials as a function of tortoise coordinate $r_{*}$ for perturbations of the scalar field on the traversable wormhole spacetime with $l=1.03$; (c)The effective potentials as a function of tortoise coordinate $r_{*}$ for perturbations of the scalar field on the traversable wormhole spacetime with $l=1.2$; (d)The effective potentials as a function of tortoise coordinate $r_{*}$ for perturbations of the scalar field on the traversable wormhole spacetime with $l=1.7$. In both cases we take $m=0.5$, $Q=0.03$ and $l_{0}=1$.}
\label{fig2}
\end{figure}

\begin{figure}[htbp]
\begin{tabular}{cc}
\begin{minipage}[t]{0.45\linewidth}
\centerline{\includegraphics[width=6.0cm]{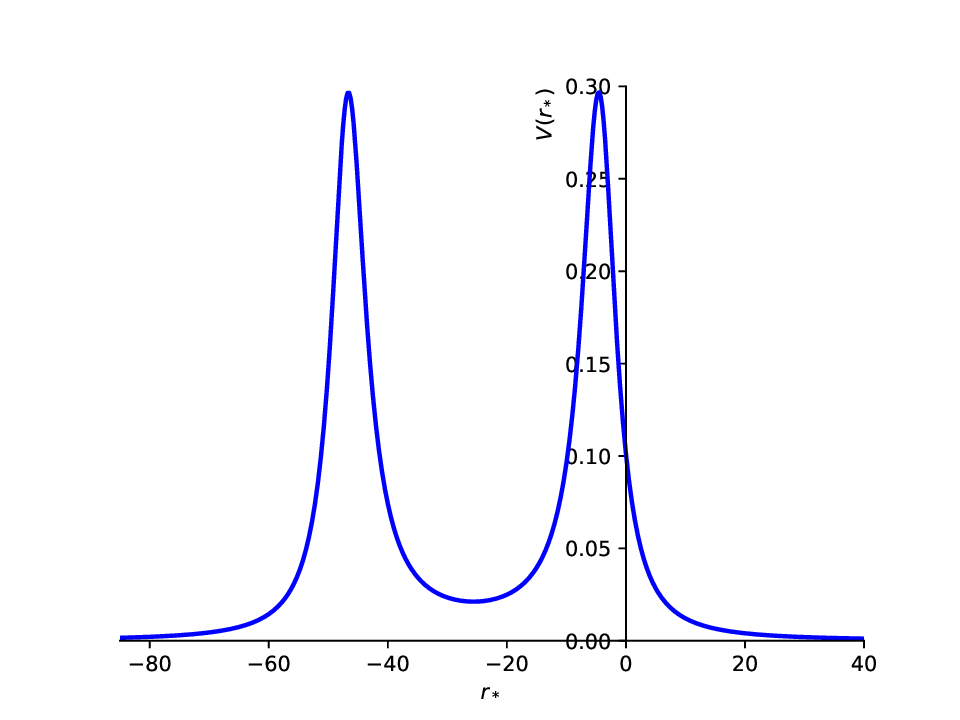}}
\centerline{(a)}
\end{minipage}
\hspace{7mm}
\begin{minipage}[t]{0.45\linewidth}
\centerline{\includegraphics[width=6.0cm]{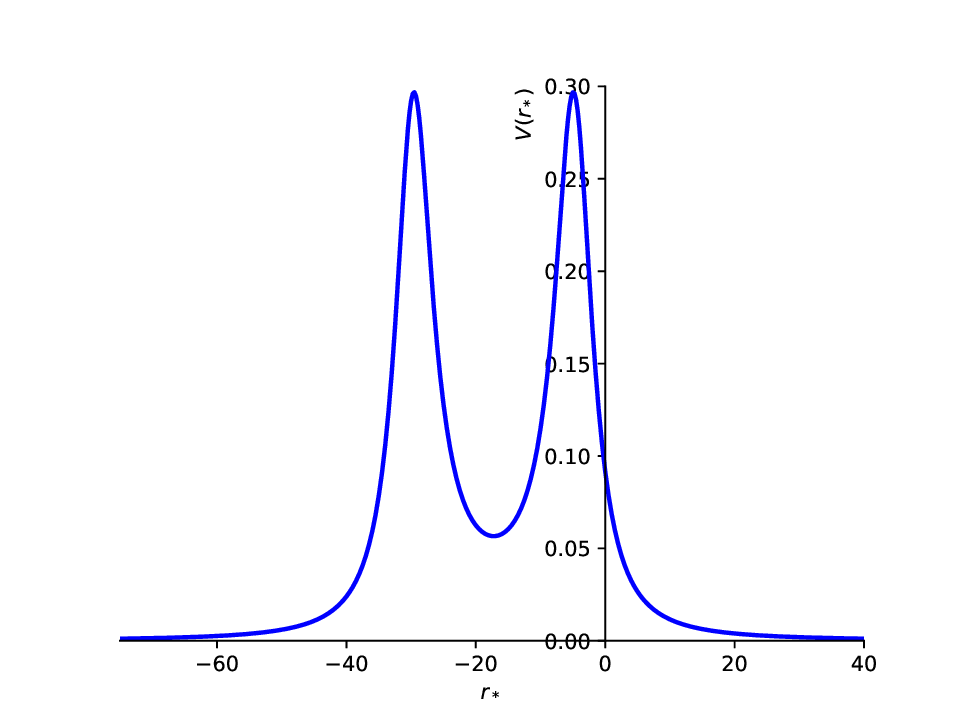}}
\centerline{(b)}
\end{minipage}
\\
\hspace{7mm}
\begin{minipage}[t]{0.45\linewidth}
\centerline{\includegraphics[width=6.0cm]{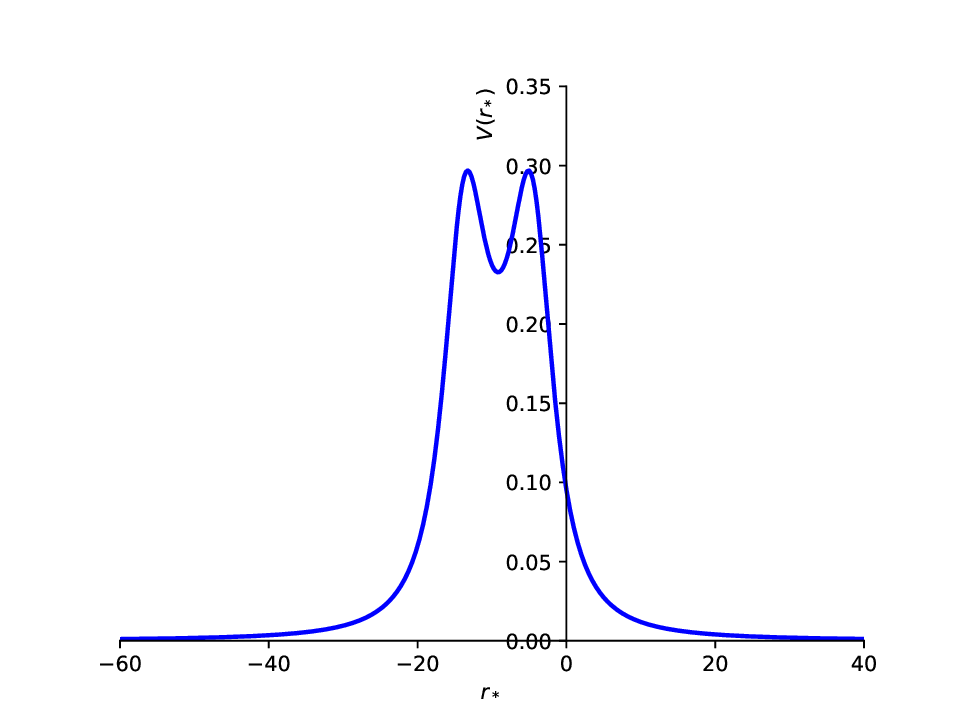}}
\centerline{(c)}
\end{minipage}

\hspace{7mm}
\begin{minipage}[t]{0.45\linewidth}
\centerline{\includegraphics[width=6.0cm]{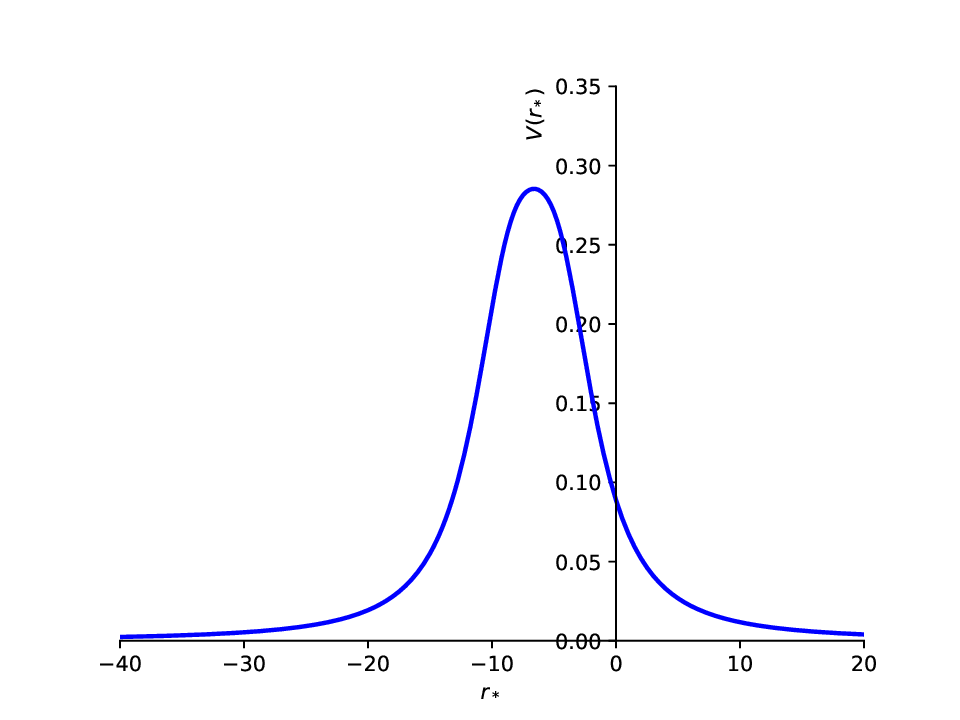}}
\centerline{(d)}
\end{minipage}
\end{tabular}
\caption{(a)The effective potential($V_{1}$) as a function of tortoise coordinate $r_{*}$ for perturbations of the electromagnetic field on the traversable wormhole spacetime with $l=1.01$; (b)The effective potential($V_{1}$) as a function of tortoise coordinate $r_{*}$ for perturbations of the electromagnetic field on the traversable wormhole spacetime with $l=1.03$; (c)The effective potential($V_{1}$) as a function of tortoise coordinate $r_{*}$ for perturbations of the electromagnetic field on the traversable wormhole spacetime with $l=1.2$; (d)The effective potential($V_{1}$) as a function of tortoise coordinate $r_{*}$ for perturbations of the electromagnetic field on the traversable wormhole spacetime with $l=1.7$. In both cases we take $m=0.5$, $Q=0.03$ and $l_{0}=1$.}
\label{fig3}
\end{figure}

\begin{figure}[htbp]
\begin{tabular}{cc}
\begin{minipage}[t]{0.45\linewidth}
\centerline{\includegraphics[width=6.0cm]{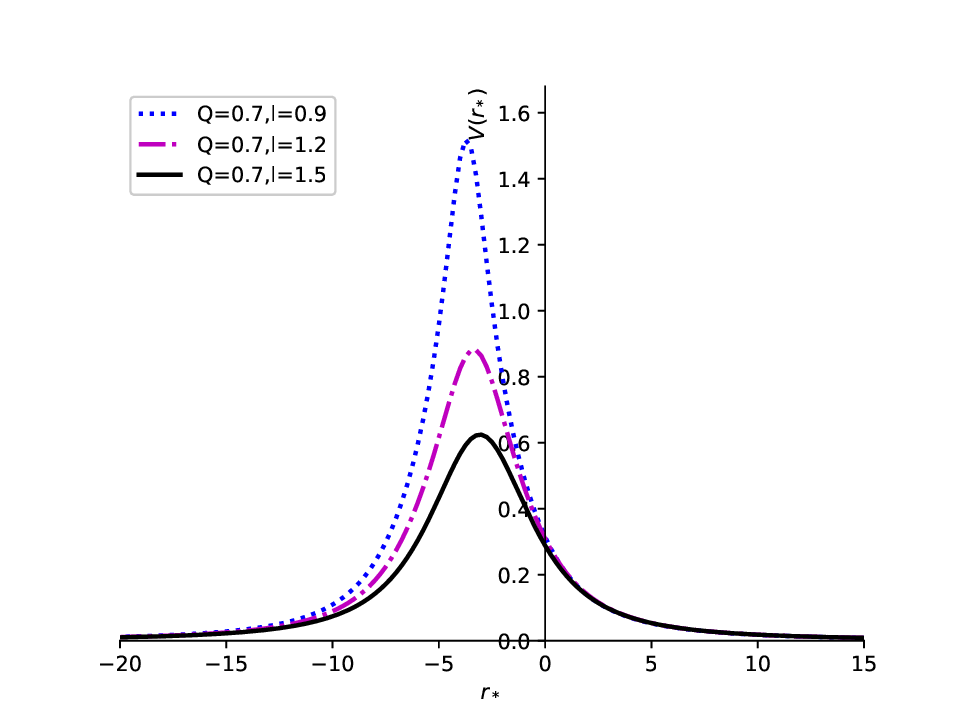}}
\centerline{(a)}
\end{minipage}
\hspace{7mm}
\begin{minipage}[t]{0.45\linewidth}
\centerline{\includegraphics[width=6.0cm]{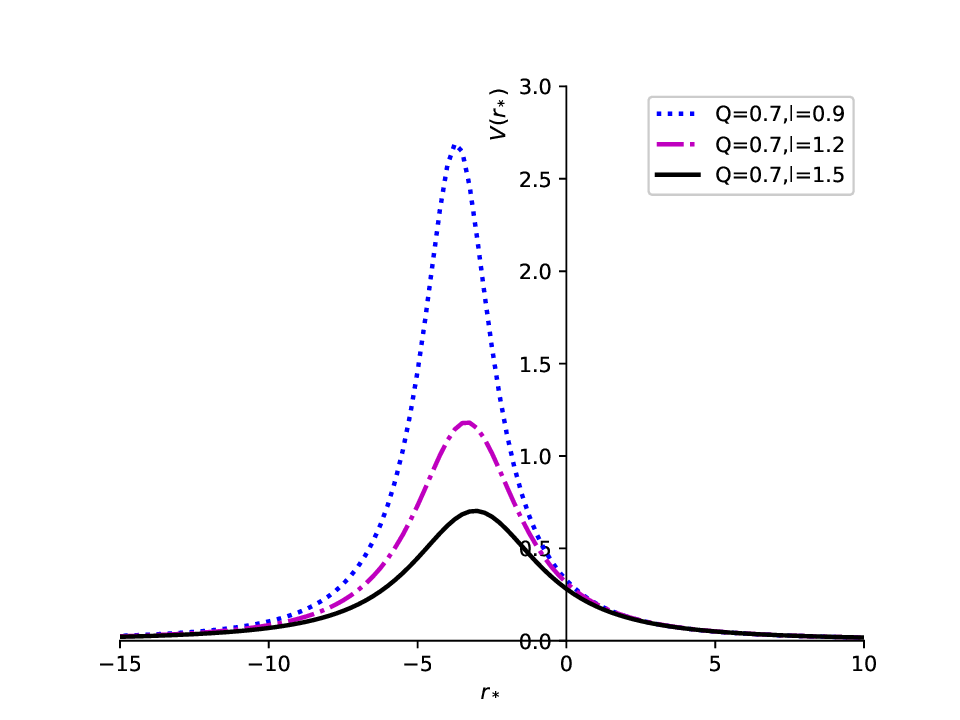}}
\centerline{(b)}
\end{minipage}
\end{tabular}
\caption{(a)The effective potentials as a function of tortoise coordinate $r_{*}$ for perturbations of the scalar field on the two-way traversable wormhole spacetime with $m=0.5$, $l_{0}=1$; (b)The effective potential($V_{1}$) as a function of tortoise coordinate $r_{*}$ for perturbations of the electromagnetic field on the two-way traversable wormhole spacetime with $m=0.5$, $l_{0}=1$.}
\label{fig4}
\end{figure}

Now, we look for the stationary solutions by using the ansatz $\Phi_{b} \sim \Psi_{b}e^{-i{\omega}t}$, thus the Regge Wheeler wave-like equation transforms into
\begin{equation}\frac{d^{2}\Psi_b (r_{*}) }{dr^{2}_{*}}+(\omega^{2}-V_b(r))\Psi_b(r_{*})=0. \label{eq31} \end{equation}
Before solving Eq.\eqref{eq31}, we must state the boundary conditions. Considering Fig.\ref{fig1}-Fig.\ref{fig4},
we have $r_{*}$ $\rightarrow$ $\pm\infty$ and $V$ $\rightarrow$ 0, thus according to the asymptotic behavior of the effective potential at the boundary, the solution of the wave function at the boundary should have the form of a plane wave, i.e. $\Psi_b(r_{*}) \sim e^{\pm i{\omega}r_{*}}$, $r_{*}$ $\rightarrow$ $+\infty$ and $\Psi_b(r_{*}) \sim e^{\pm i{\omega}r_{*}}$, $r_{*}$ $\rightarrow$ $-\infty$. Since we are dealing with the black bounce Reissner Nordstr{\"o}m geometry, thus we need to study based on the black hole and the wormhole. For the case of the black hole, the quasinormal boundary conditions imply pure incoming waves at the horizon and pure outgoing waves at spatial infinity, thus for the asymptotically flat solutions, the quasinormal boundary conditions are \cite{ret37}
\begin{equation} \Psi_b \propto e^{\pm i{\omega}r_{*}}, r_{*} \rightarrow \pm\infty. \label{eq32} \end{equation}

For the case of the wormhole, the quasinormal boundary conditions for both infinities are the requirement of purely outgoing waves into both infinities, which implies that no waves are
coming from both asymptotically flat regions, i.e. $\Psi_b \propto e^{\pm i{\omega}r_{*}}, r_{*} \rightarrow \pm\infty$ \cite{ret38}. By comparison the black hole and the wormhole, we see that they have a qualitatively different situation, yet leading to the same boundary conditions,  thus in some sense, the throat of the wormhole plays the role of the event horizon of the black hole and we can use the same method for both cases. Moreover, the solution of a wave function satisfies the above boundary conditions, whose frequencies are a set of discrete complex numbers.

Fig.\ref{fig1}-Fig.\ref{fig4} report that the effective potentials change with different parameters. In Fig.\ref{fig1}, the value of parameters satisfy the conditions of $\rvert {Q}\rvert\le m$ and $(\rvert {l}\rvert < m- \sqrt{m ^{2}-Q^{2} })$, so Fig.\ref{fig1} is the effective potentials of regular black hole, which show that the maximum value of the effective potential reveal no significant changes when we remain the constant charge parameter $Q$ and set the changeable bounce parameter $l$. In Fig.\ref{fig2}-Fig.\ref{fig3}, the value of parameters satisfy the conditions of $(\rvert {Q}\rvert \le m)$ and $(\rvert {l}\rvert > m+ \sqrt{m ^{2}-Q^{2}})$, so Fig.\ref{fig2}-Fig.\ref{fig3} are corresponding to the traversable wormhole for perturbations of the scalar field and perturbations of the electromagnetic field respectively. For both cases, which show that the effective potential has obvious double peaks, and with the increasing parameter $l$, the double peaks will eventually merge to a single peak. While by comparing the maximum value of the effective potential for perturbations of the scalar field and perturbations of the electromagnetic field, we see that the former case is greater than the latter case for the same spacetimes parameters. In Fig.\ref{fig4}, the value of parameters satisfy the conditions of $\rvert {Q}\rvert > m$, therefore the charged black-bounce spacetimes will transform into the two-way traversable wormhole, we can see that if remaining the constant charge parameter $Q$, the maximum values of the effective potential decrease with the bounce parameter $l$.

\section{The time domain integration method}\label{sec3}
The external perturbation field satisfies the equation under the charged black-bounce spacetimes in terms of the tortoise coordinate is given by
\begin{equation}\frac{d^{2}\Phi_b}{dt^{2}}-\frac{d^{2}\Phi_b }{dr^{2}_{*}}+V_b(r)\Phi_b=0, \label{eq33} \end{equation}
since QNMs are complex frequencies related with purely outgoing waves at spatial infinity, from Eq.\eqref{eq33} we have $\Phi_b(r_{*},t)$ $\rightarrow$ $e^{\pm i{\omega}r_{*}}$$e^{-i{\omega}t}$ as $r_{*}$ $\rightarrow$ $\pm \infty$, and for an asymptotically flat spacetime, $V_{b}$ approaches 0. Based on the features of the effective potentials, we will deal with the scalar and vector QNMs in context of the charged black-bounce spacetimes, the existence of QNMs can be directly observed in the time evolution of the scalar field and vector field obtained by integrating the wave equation (Eq.\eqref{eq33}). In view of the obscure expression of the effective potentials, it is difficult to obtain the analytical solutions, here we intend to adopt the time domain integration method which does not depend on the form of the potential barrier, thus we rewritten Eq.\eqref{eq33} in terms of the light-cone variables $u=t-r_{*}$ and $v=t+r_{*}$ as
\begin{equation}-4\frac{\partial^{2}}{\partial{u}\partial{v}}\Phi_b(u,v)=V(r(u,v))\Phi_b(u,v), \label{eq34} \end{equation}
where $u$ and $v$ are integral constants.

Next, we move on towards the numerical integration of the Eq.\eqref{eq34}, regarding the characteristic initial value problem, the initial data is specified on the two null surfaces $u = u_0$ and $v = v_0$. In view of the fact that the field decay is almost independent of the initial conditions, and the initial condition is the Gaussian wave packet \cite{ret39,ret40,ret41}, thus we start with a Gaussian pulse ($u = u_0$) and set the field to zero ($v = v_0$), i.e.
\begin{equation}\Phi_b(u=u_0,v)=exp[-\frac{(v-v_c)^{2}}{2{\sigma}^{2}}], \Phi_b(u,v = v_0)=0,  \label{eq35} \end{equation}
where the Gaussian pulse with a width is $\sigma=3$ and the center of the Gaussian pulse is $v_c=10$. In addition, the appropriate discretization scheme in terms of the Taylor¡¯s theorem \cite{ret42,ret43} is
\begin{equation}\begin{aligned} \Phi_b(N)= &\Phi_b(w)+\Phi_b(E)-\Phi_b(S)-\\ &\Delta^{2}\frac{V(w)\Phi_b(w)+V(E)\Phi_b(E)} {8}+O(\Delta^{4}), \end{aligned} \label{eq36} \end{equation}
where the following designations for the points were used: $N=(u+\Delta,v+\Delta)$, $w=(u+\Delta,v)$, $E=(u,v+\Delta)$ and $S=(u,v)$, thus using the corresponding initial conditions along u and v lines we numerically integrate to derive the time-domain profiles. In addition, in order to facilitate the extraction of the quasinormal frequencies, we will employ the Prony method\cite{ret44} which helps us to fit the signal by a sum of exponents with some excitation factors.

\section{The pictures of echoes for the charged black-bounce spacetimes}\label{sec4}
In this section, we concentrate on the time evolution under the perturbations of massless scalar field and electromagnetic field in the charged black-bounce spacetimes. Specifically, we depict the time evolution which has the same parameters with the effective potential in order to facilitate the control study, and it should be noted that the necessary condition for the appearance of the echoes is that the potential well must appear.

\begin{figure}[htbp]
\begin{tabular}{cc}
\begin{minipage}[t]{0.45\linewidth}
\centerline{\includegraphics[width=6.0cm]{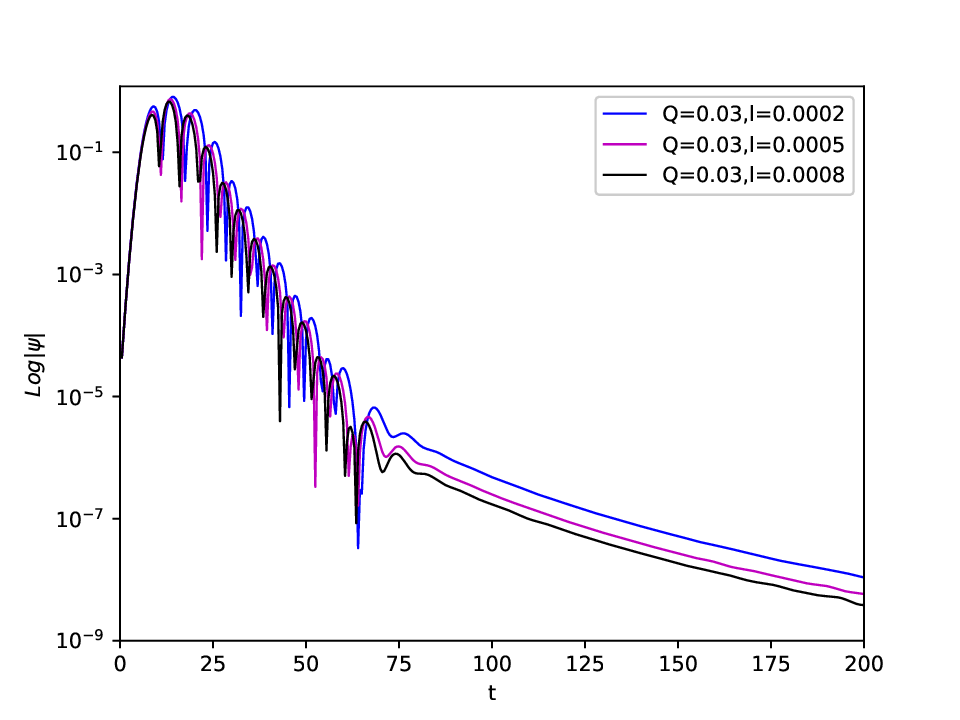}}
\centerline{(a)}
\end{minipage}
\hspace{7mm}
\begin{minipage}[t]{0.45\linewidth}
\centerline{\includegraphics[width=6.0cm]{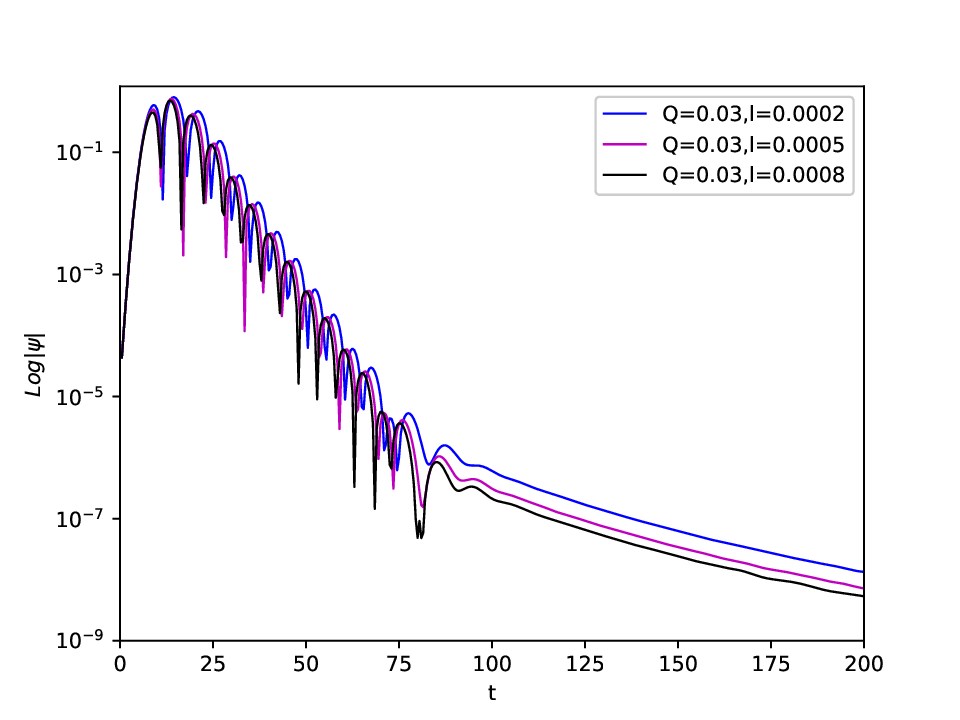}}
\centerline{(b)}
\end{minipage}
\end{tabular}
\caption{(a)The time evolution of regular black hole for perturbations of the scalar field with $m=0.5$, $l_{0}=1$; (b)The time evolution of regular black hole for perturbations of the electromagnetic field($V_{1}$) with $m=0.5$, $l_{0}=1$.}
\label{fig5}
\end{figure}

\begin{figure}[htbp]
\begin{tabular}{cc}
\begin{minipage}[t]{0.45\linewidth}
\centerline{\includegraphics[width=6.0cm]{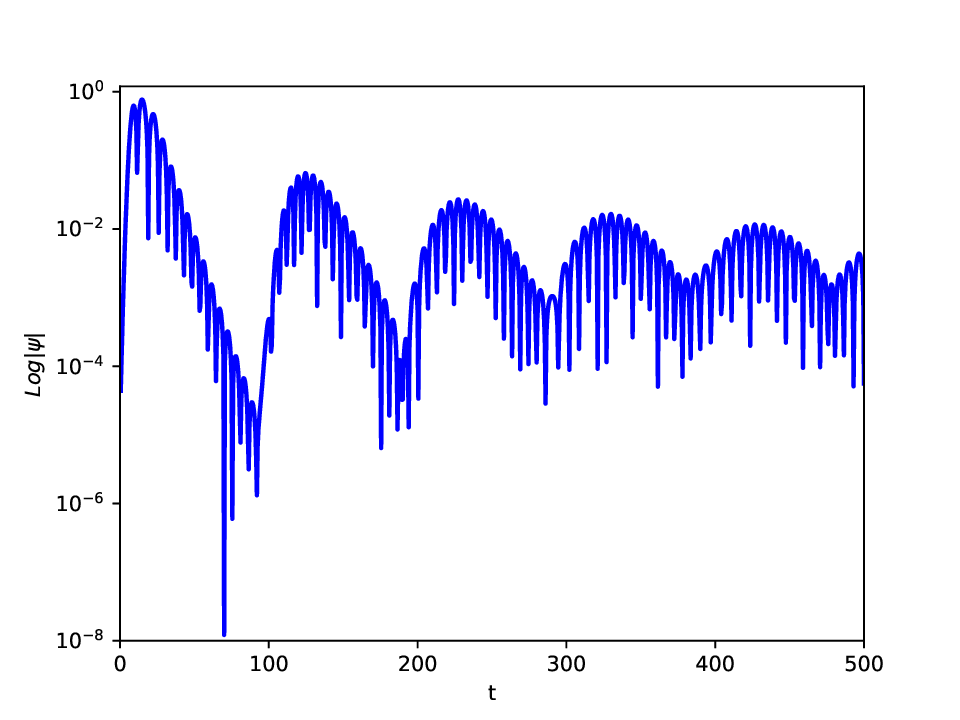}}
\centerline{(a)}
\end{minipage}
\hspace{7mm}
\begin{minipage}[t]{0.45\linewidth}
\centerline{\includegraphics[width=6.0cm]{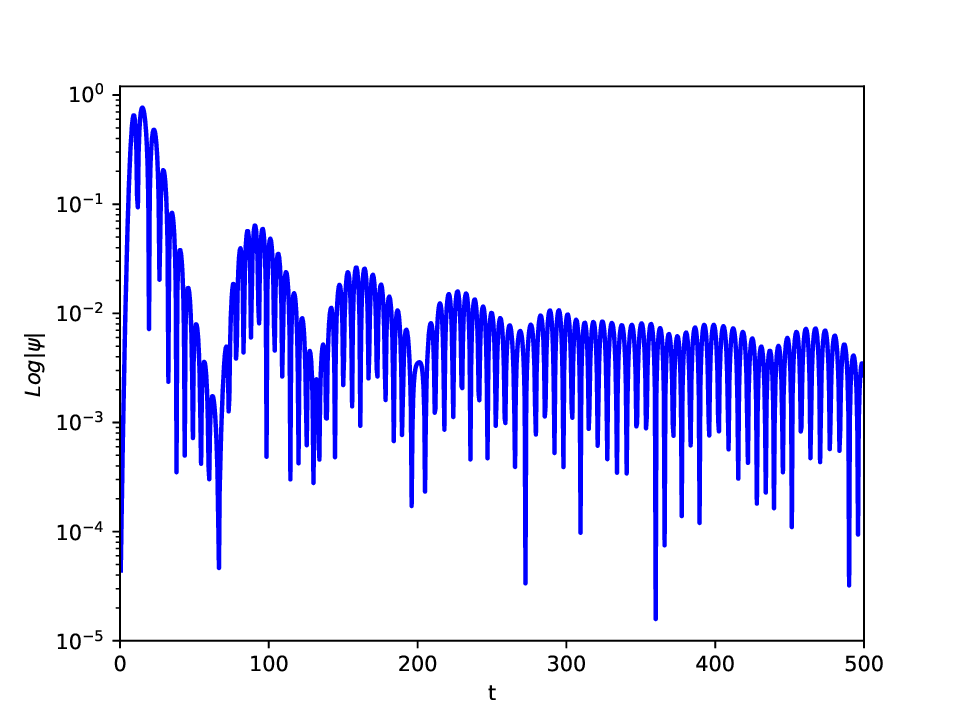}}
\centerline{(b)}
\end{minipage}
\hspace{7mm}
\\
\begin{minipage}[t]{0.45\linewidth}
\centerline{\includegraphics[width=6.0cm]{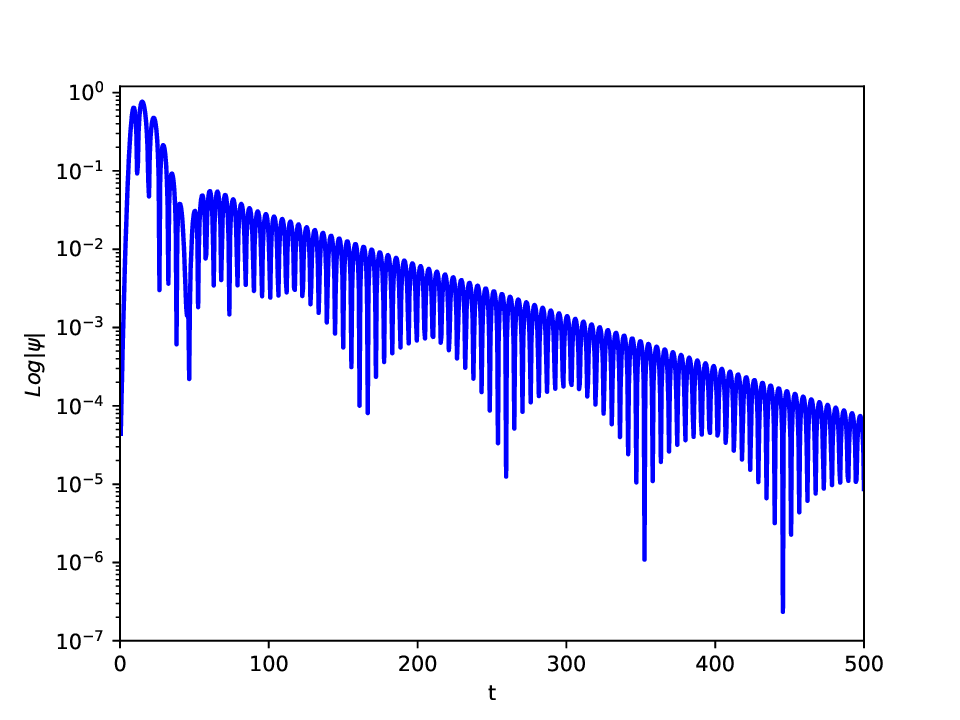}}
\centerline{(c)}
\end{minipage}

\hspace{7mm}
\begin{minipage}[t]{0.45\linewidth}
\centerline{\includegraphics[width=6.0cm]{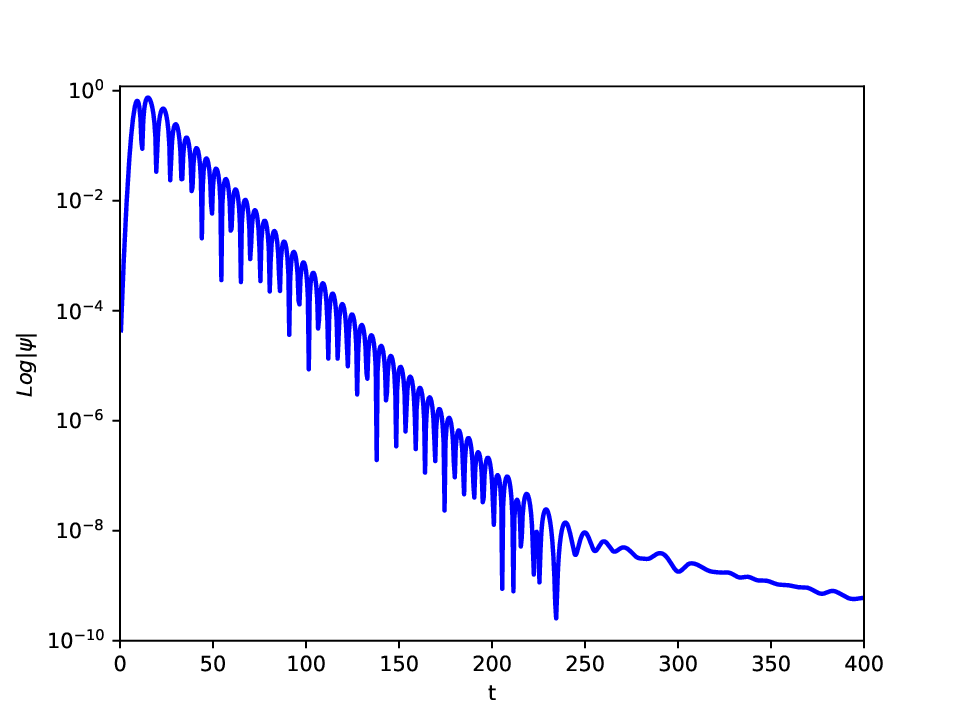}}
\centerline{(d)}
\end{minipage}
\end{tabular}
\caption{(a)The time evolution of traversable wormhole for perturbations of the scalar field with $l=1.01$; (b)The time evolution of traversable wormhole for perturbations of the scalar field with $l=1.03$; (c)The time evolution of traversable wormhole for perturbations of the scalar field with $l=1.2$; (d)The time evolution of traversable wormhole for perturbations of the scalar field with $l=1.7$. In both cases we take $m=0.5$, $Q=0.03$ and $l_{0}=1$.}
\label{fig6}
\end{figure}

\begin{figure}[htbp]
\begin{tabular}{cc}
\begin{minipage}[t]{0.45\linewidth}
\centerline{\includegraphics[width=6.0cm]{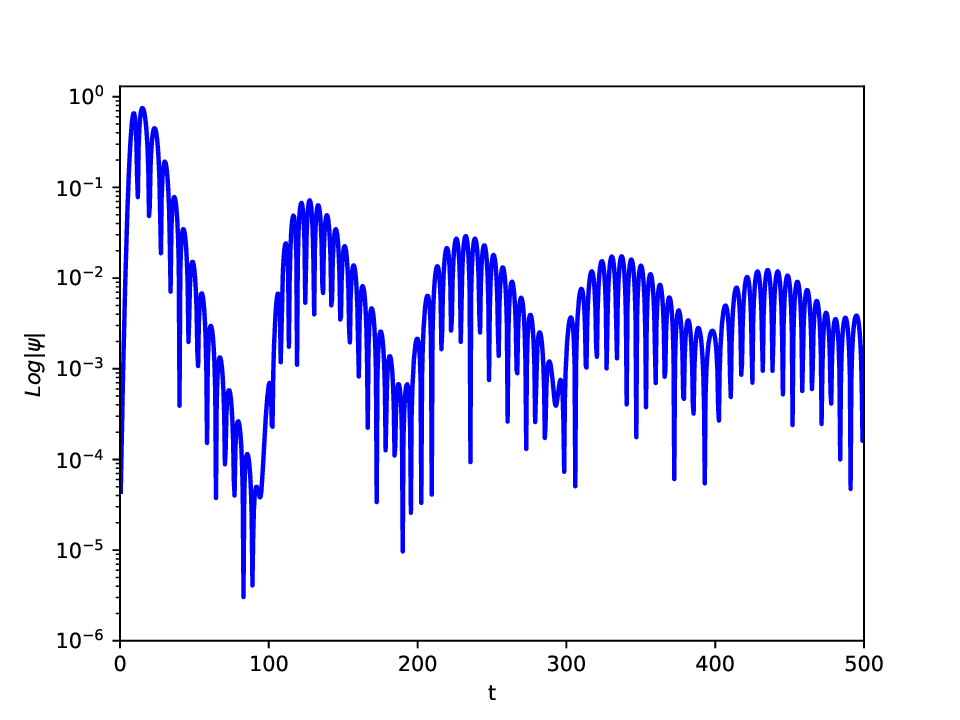}}
\centerline{(a)}
\end{minipage}
\hspace{7mm}
\begin{minipage}[t]{0.45\linewidth}
\centerline{\includegraphics[width=6.0cm]{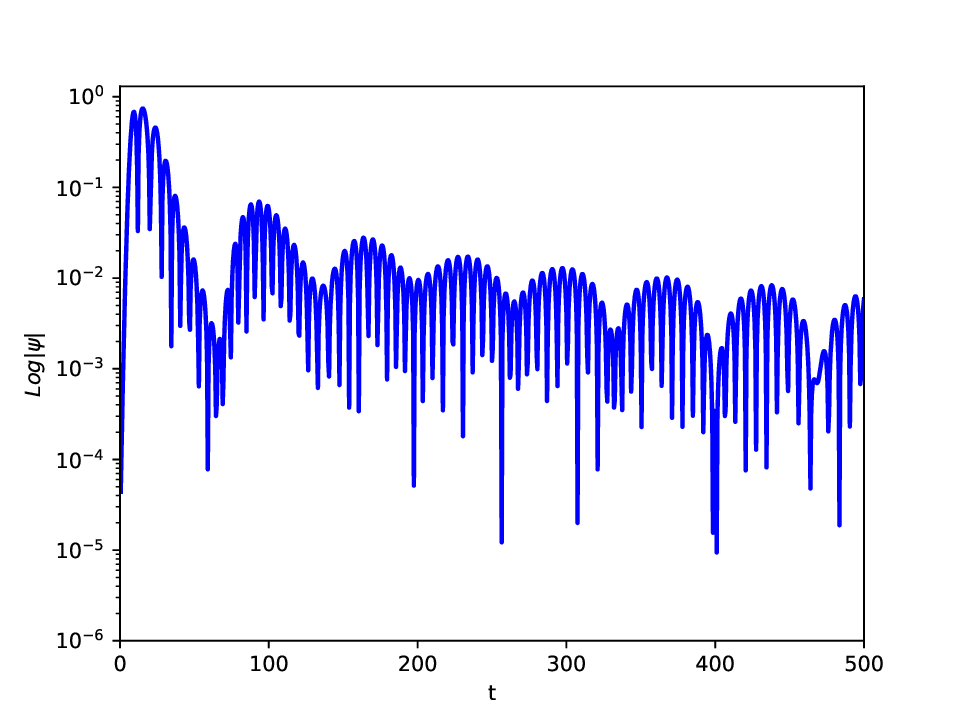}}
\centerline{(b)}
\end{minipage}
\\
\hspace{7mm}
\begin{minipage}[t]{0.45\linewidth}
\centerline{\includegraphics[width=6.0cm]{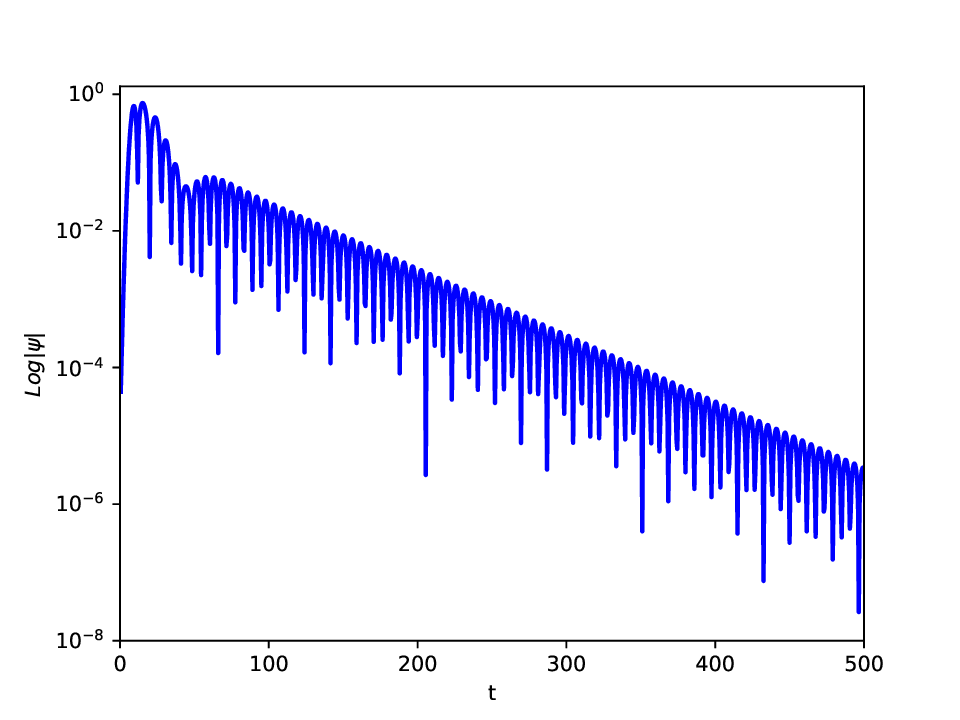}}
\centerline{(c)}
\end{minipage}

\hspace{7mm}
\begin{minipage}[t]{0.45\linewidth}
\centerline{\includegraphics[width=6.0cm]{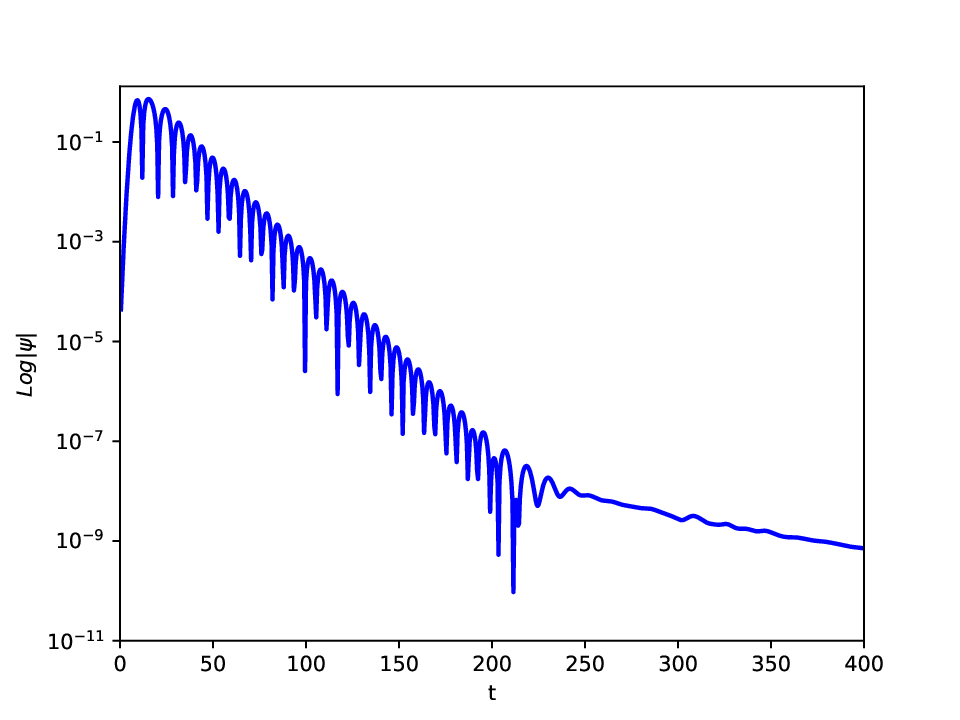}}
\centerline{(d)}
\end{minipage}
\end{tabular}
\caption{(a)The time evolution of traversable wormhole for perturbations of the electromagnetic field($V_{1}$) with $l=1.01$; (b)The time evolution of traversable wormhole for perturbations of the electromagnetic field($V_{1}$) with $l=1.03$; (c)The time evolution of traversable wormhole for perturbations of the electromagnetic field($V_{1}$) with $l=1.2$; (d)The time evolution of traversable wormhole for perturbations of the electromagnetic field($V_{1}$) with $l=1.7$. In both cases we take $m=0.5$, $Q=0.03$ and $l_{0}=1$.}
\label{fig7}
\end{figure}

\begin{figure}[htbp]
\begin{tabular}{cc}
\begin{minipage}[t]{0.45\linewidth}
\centerline{\includegraphics[width=6.0cm]{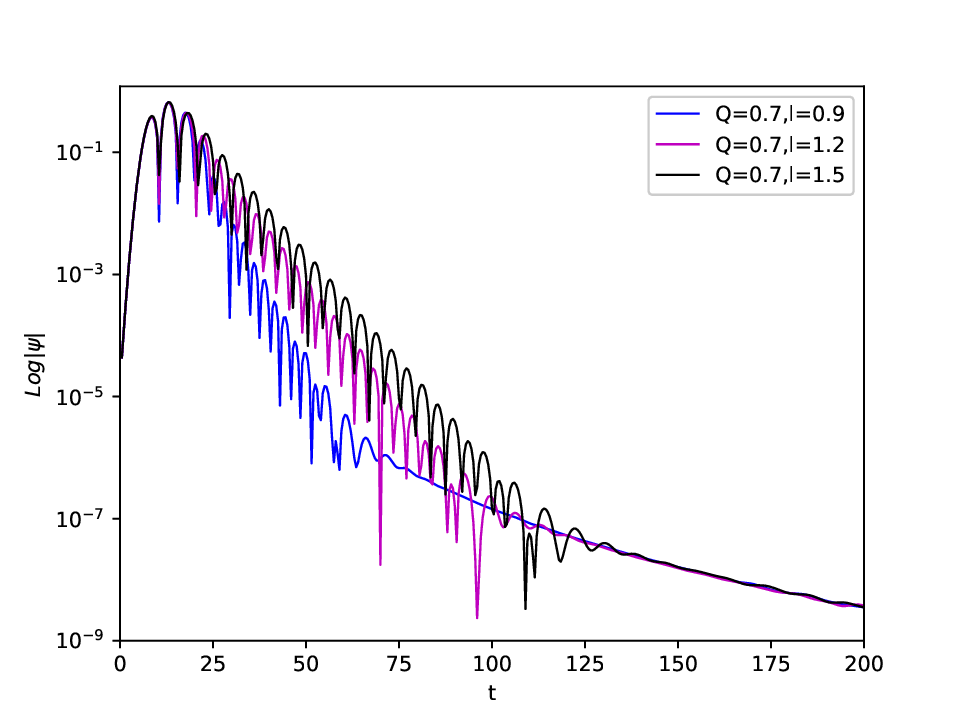}}
\centerline{(a)}
\end{minipage}
\hspace{7mm}
\begin{minipage}[t]{0.45\linewidth}
\centerline{\includegraphics[width=6.0cm]{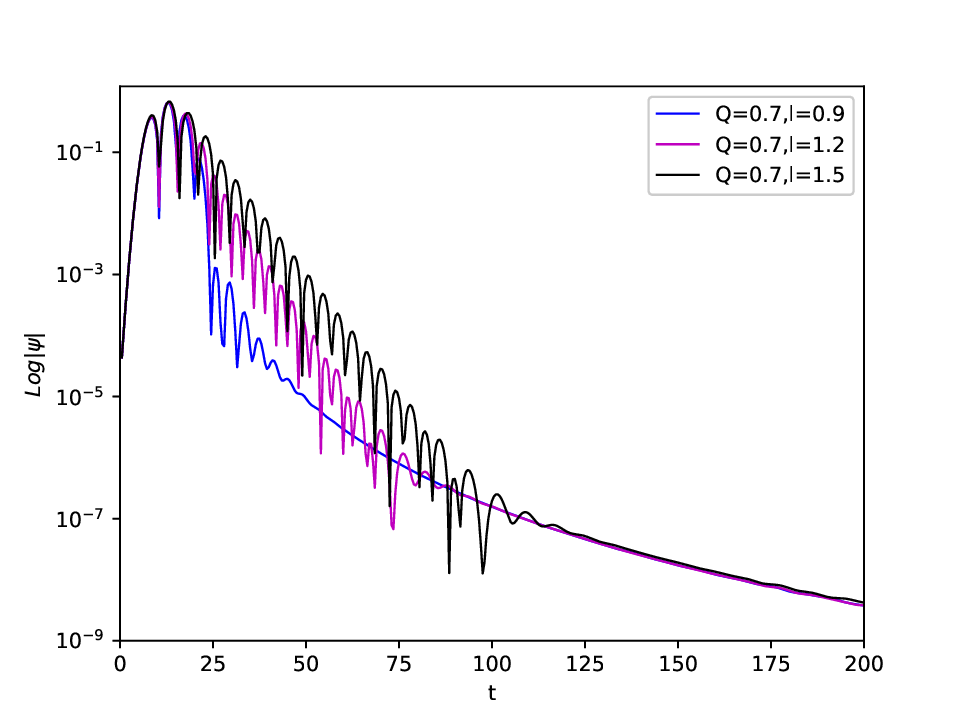}}
\centerline{(b)}
\end{minipage}
\end{tabular}
\caption{(a)The time evolution of two-way traversable wormhole  for perturbations of the scalar field with $m=0.5$, $l_{0}=1$; (b)The time evolution of two-way traversable wormhole  for perturbations of the electromagnetic field($V_{1}$) with $m=0.5$, $l_{0}=1$.}
\label{fig8}
\end{figure}

\begin{figure}[htbp]
\begin{tabular}{cc}
\begin{minipage}[t]{0.45\linewidth}
\centerline{\includegraphics[width=6.0cm]{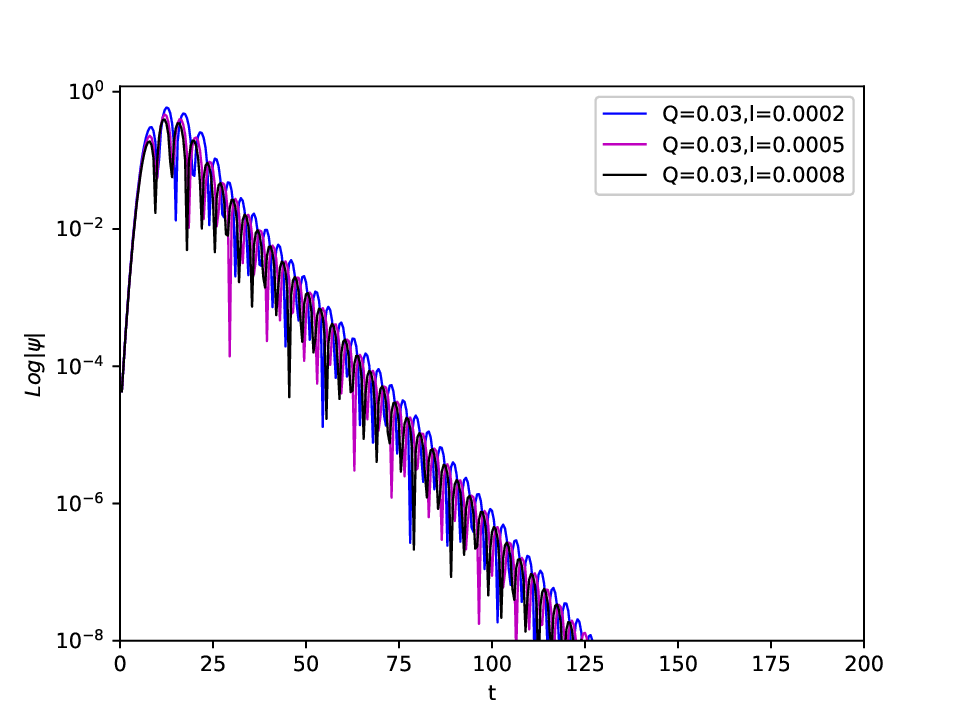}}
\centerline{(a)}
\end{minipage}
\hspace{7mm}
\begin{minipage}[t]{0.45\linewidth}
\centerline{\includegraphics[width=6.0cm]{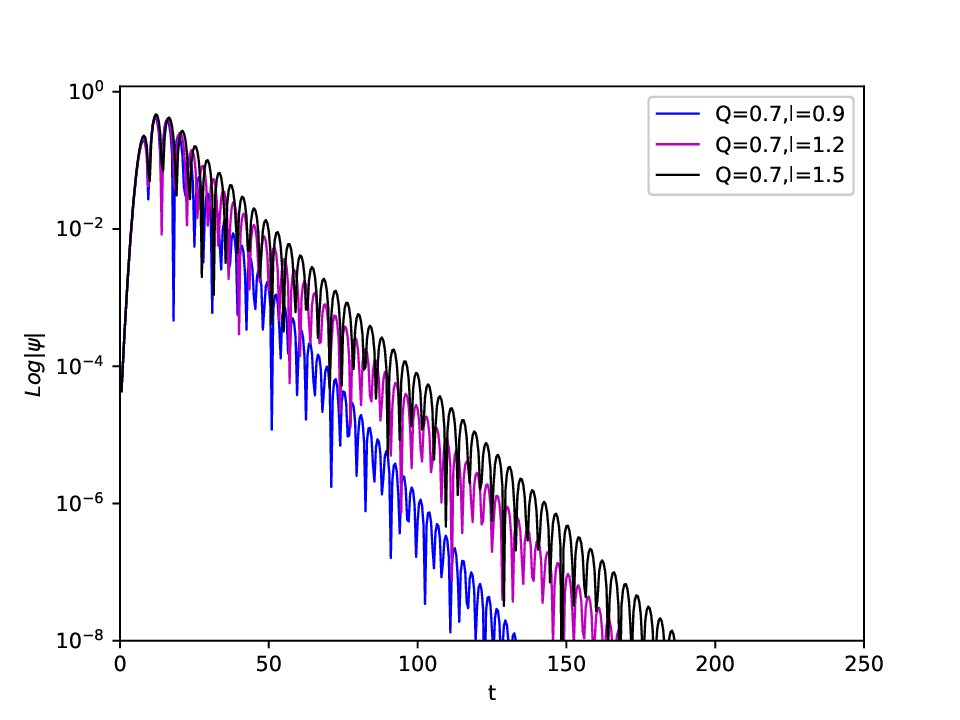}}
\centerline{(b)}
\end{minipage}
\end{tabular}
\caption{(a)The time evolution of regular black hole for perturbations of the electromagnetic field($V_{2}$) with $m=0.5$, $l_{0}=2$. (b)The time evolution of two-way traversable wormhole  for perturbations of the electromagnetic field($V_{2}$) with $m=0.5$, $l_{0}=2$.}
\label{fig9}
\end{figure}

\begin{figure}[htbp]
\begin{tabular}{cc}
\begin{minipage}[t]{0.45\linewidth}
\centerline{\includegraphics[width=6.0cm]{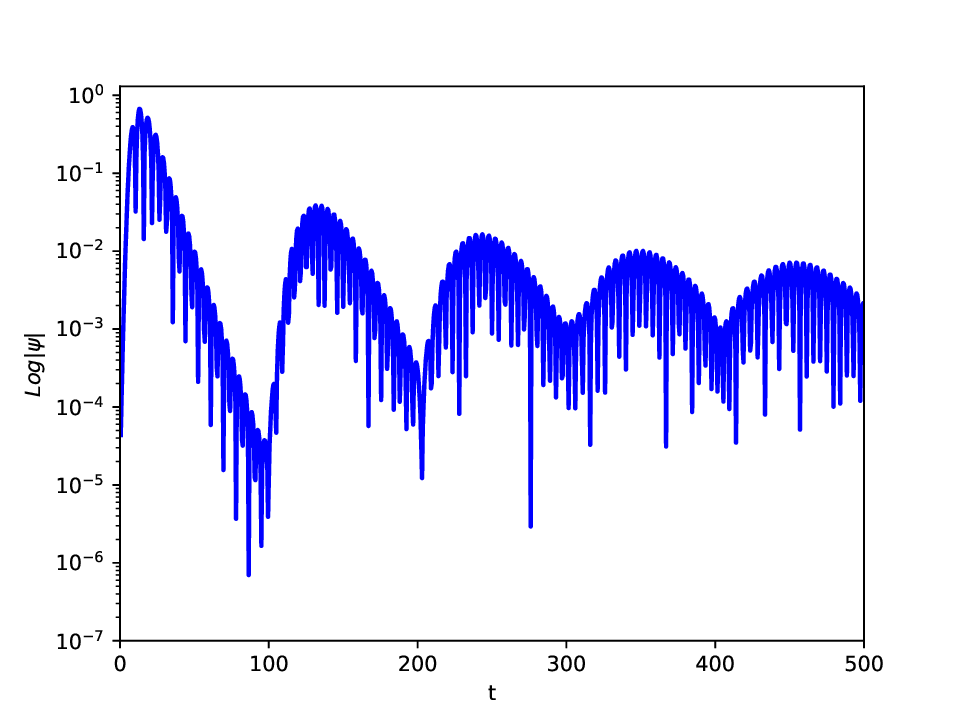}}
\centerline{(a)}
\end{minipage}
\hspace{7mm}
\begin{minipage}[t]{0.45\linewidth}
\centerline{\includegraphics[width=6.0cm]{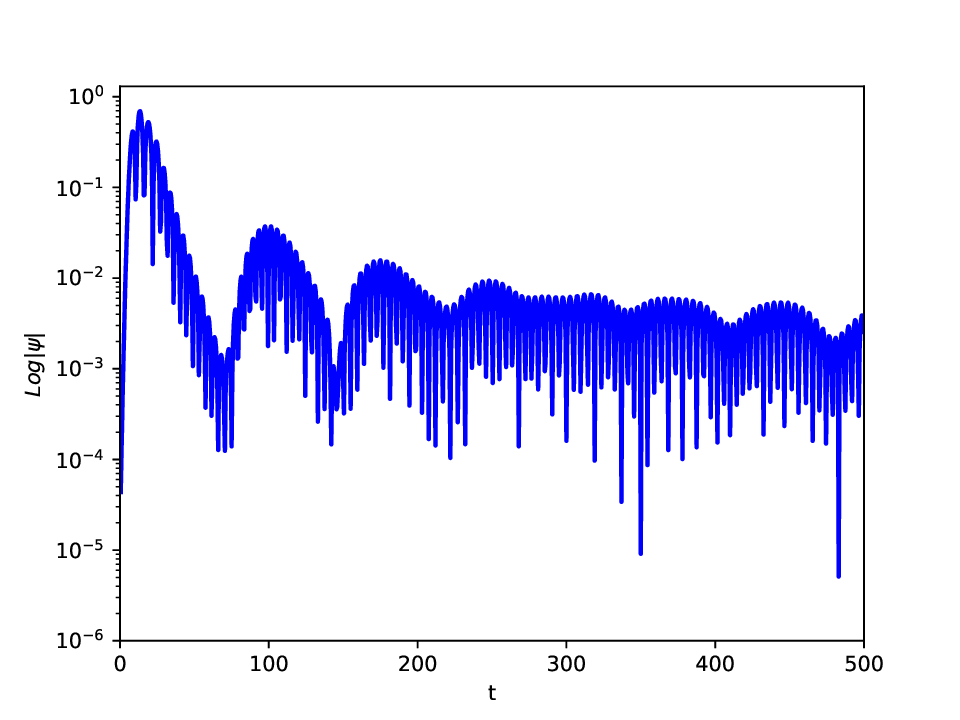}}
\centerline{(b)}
\end{minipage}
\\
\hspace{7mm}
\begin{minipage}[t]{0.45\linewidth}
\centerline{\includegraphics[width=6.0cm]{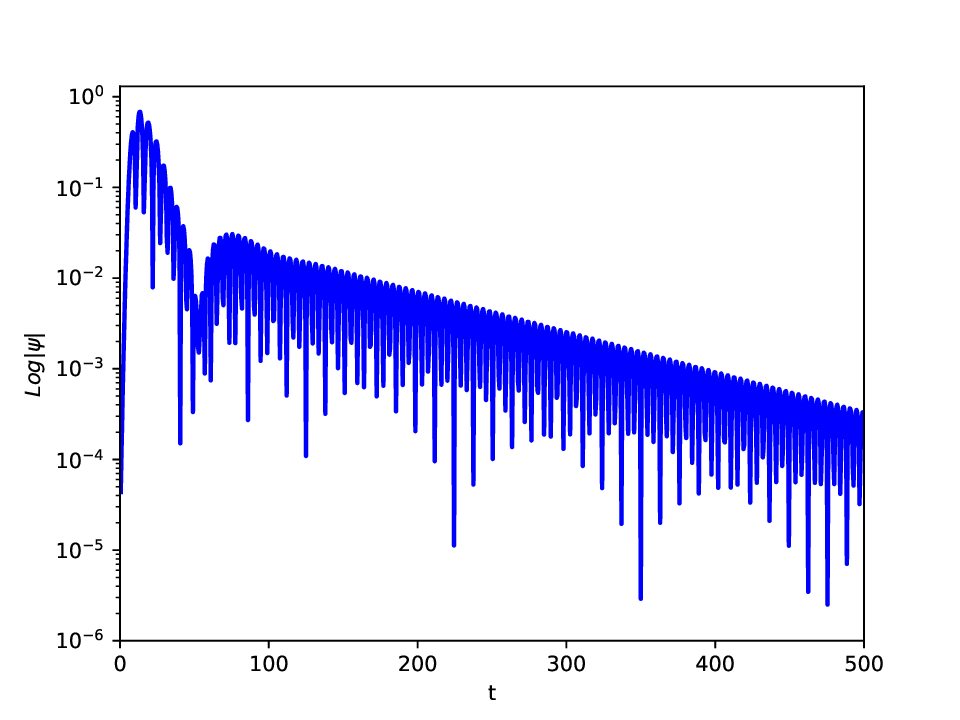}}
\centerline{(c)}
\end{minipage}

\hspace{7mm}
\begin{minipage}[t]{0.45\linewidth}
\centerline{\includegraphics[width=6.0cm]{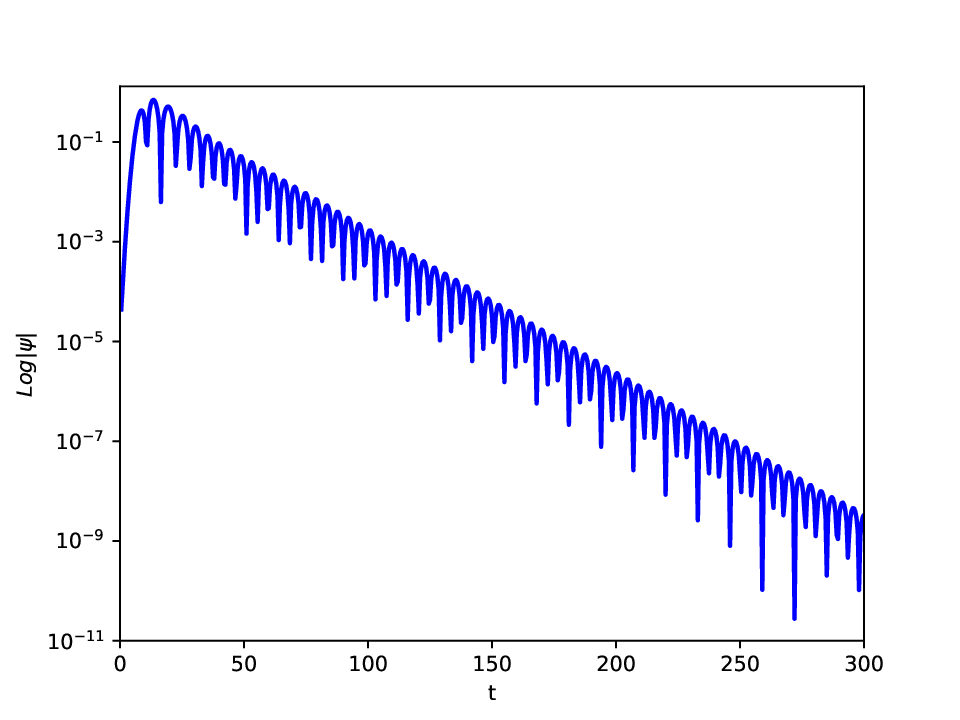}}
\centerline{(d)}
\end{minipage}
\end{tabular}
\caption{(a)The time evolution of traversable wormhole for perturbations of the electromagnetic field($V_{2}$) with $l=1.01$; (b)The time evolution of traversable wormhole for perturbations of the electromagnetic field($V_{2}$) with $l=1.03$; (c)The time evolution of traversable wormhole for perturbations of the electromagnetic field($V_{2}$) with $l=1.2$; (d)The time evolution of traversable wormhole for perturbations of the electromagnetic field($V_{2}$) with $l=1.7$. In both cases we take $m=0.5$, $Q=0.03$ and $l_{0}=2$.}
\label{fig10}
\end{figure}

\begin{figure}[htbp]
\begin{tabular}{cc}
\begin{minipage}[t]{0.45\linewidth}
\centerline{\includegraphics[width=6.0cm]{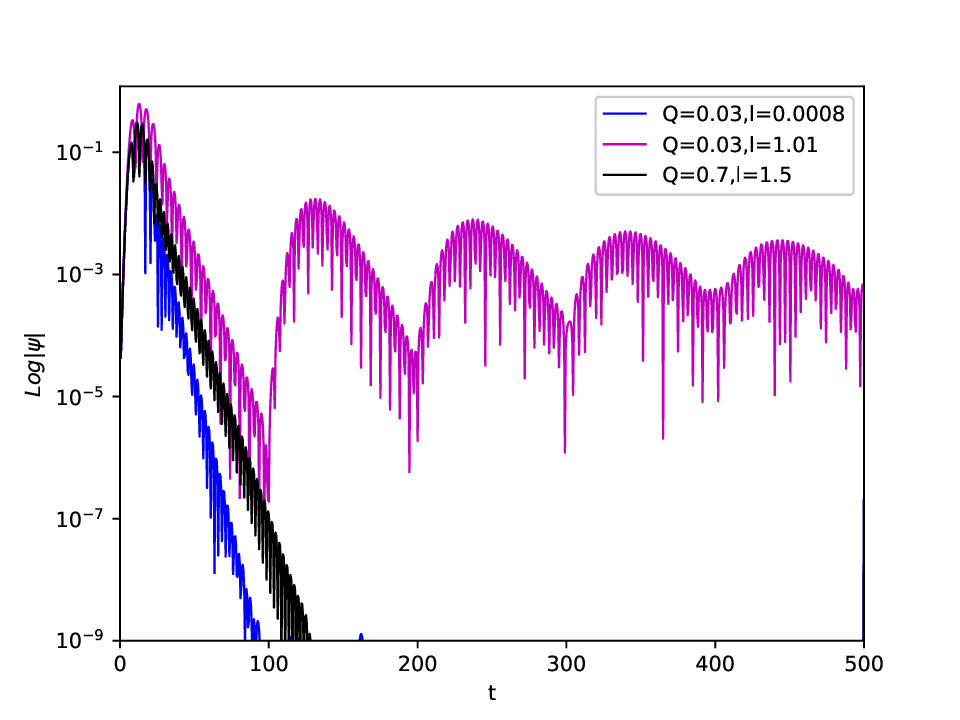}}
\centerline{(a)}
\end{minipage}
\hspace{7mm}
\begin{minipage}[t]{0.45\linewidth}
\centerline{\includegraphics[width=6.0cm]{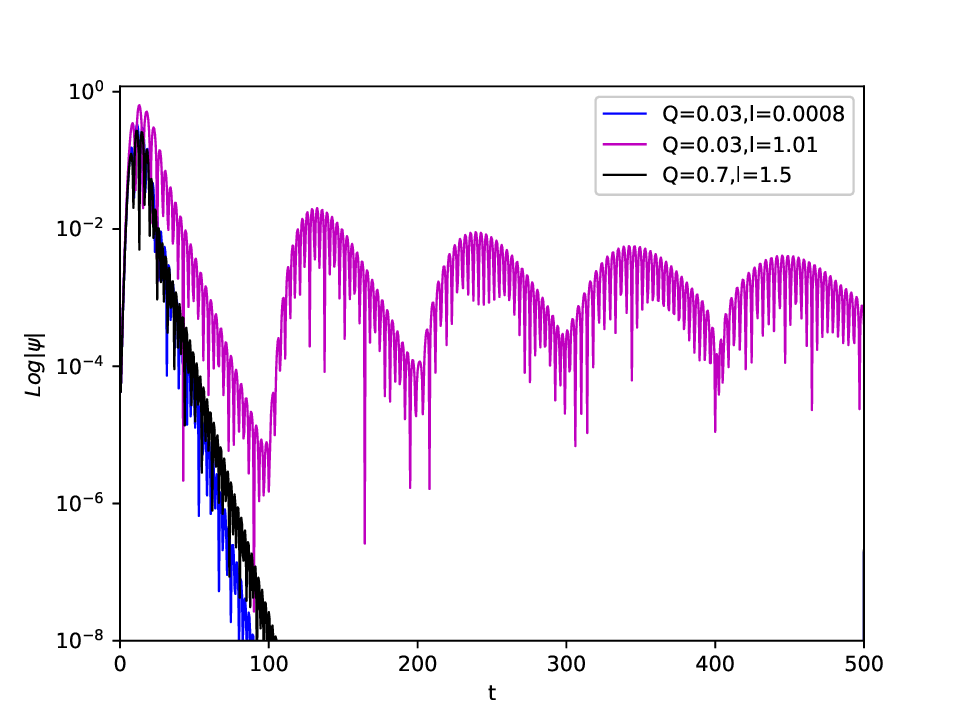}}
\centerline{(b)}
\end{minipage}
\end{tabular}
\caption{(a)Time-domain profiles for scalar field in charged black-bounce spacetimes; (b)Time-domain profiles for electromagnetic field($V_{1}$) in charged black-bounce spacetimes; In both cases we take $m=0.5$ and $l_{0}=2$.}
\label{fig11}
\end{figure}

\begin{figure}[htbp]
\begin{tabular}{cc}
\begin{minipage}[t]{0.45\linewidth}
\centerline{\includegraphics[width=6.0cm]{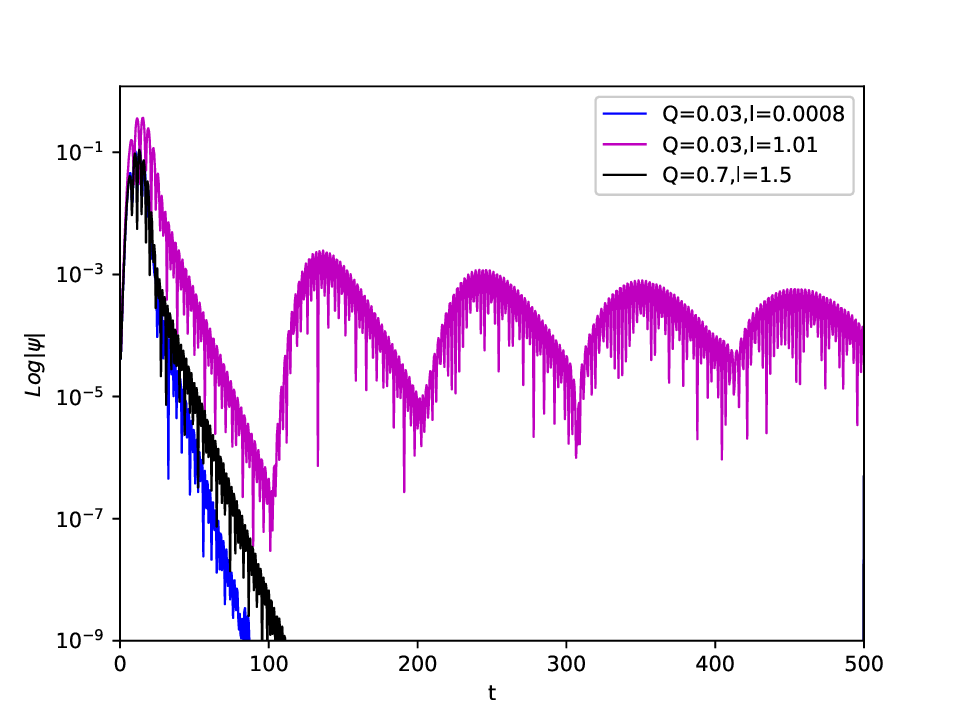}}
\centerline{(a)}
\end{minipage}
\hspace{7mm}
\begin{minipage}[t]{0.45\linewidth}
\centerline{\includegraphics[width=6.0cm]{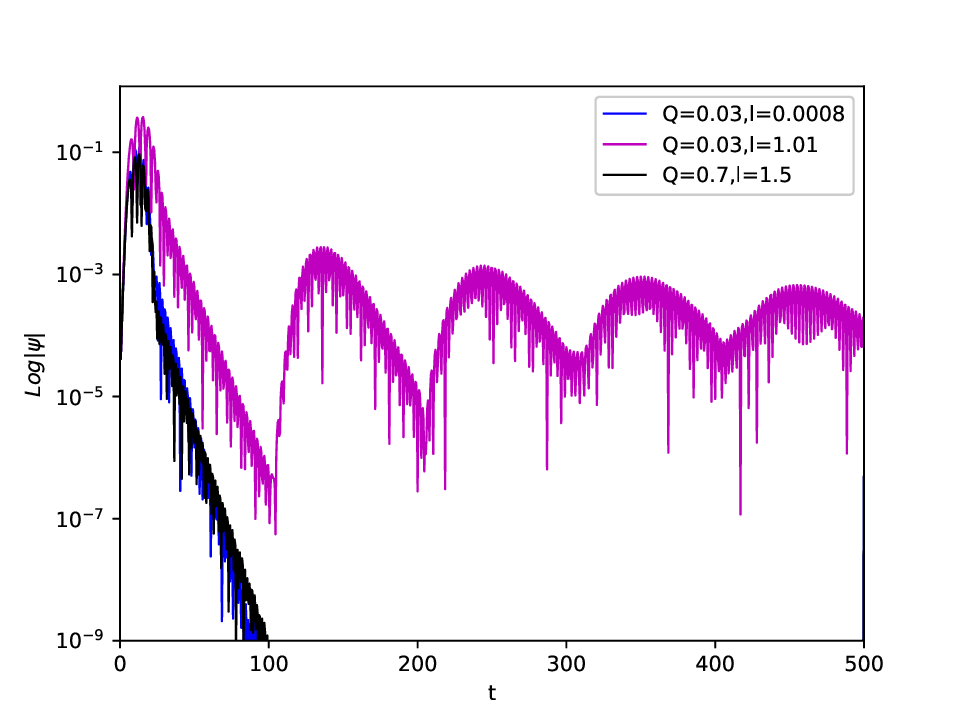}}
\centerline{(b)}
\end{minipage}
\\
\hspace{7mm}
\begin{minipage}[t]{0.45\linewidth}
\centerline{\includegraphics[width=6.0cm]{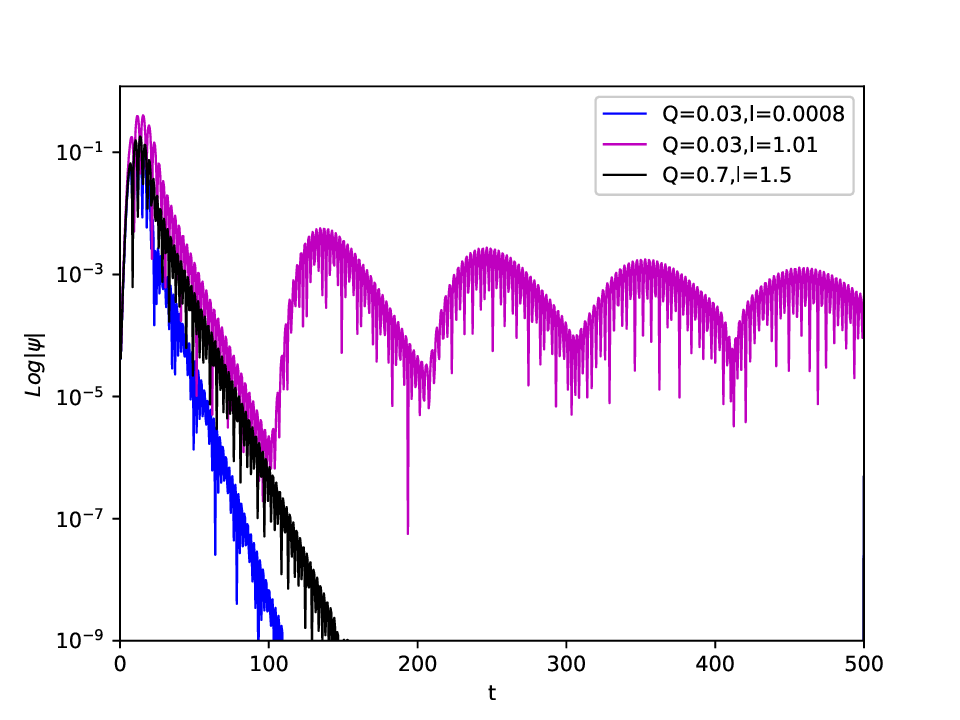}}
\centerline{(c)}
\end{minipage}

\end{tabular}
\caption{(a)Time-domain profiles for scalar field in charged black-bounce spacetimes; (b)Time-domain profiles for electromagnetic field($V_{1}$) in charged black-bounce spacetimes; (c)Time-domain profiles for electromagnetic field($V_{2}$) in charged black-bounce spacetimes; In both cases we take $m=0.5$ and $l_{0}=3$. }
\label{fig12}
\end{figure}

Fig.\ref{fig5} expresses the time evolution of regular black hole ($\rvert {Q}\rvert \le m$,$\rvert {l}\rvert < m- \sqrt{m ^{2}-Q^{2} }$) for perturbations of the scalar field and  perturbations of the electromagnetic field, which shows that there are three stages instead of echo signal, moreover, we find that the quasinormal ringdown of regular black hole will decay slighter slower with the decreasing bounce parameter $l$. Actually, this result can be predicted from Fig.\ref{fig1}, since which reveals no potential well in the effective potential. While for the same parameters, the decay rate of the quasinormal ringdown for perturbations of the scalar field is faster than the decay rate of the quasinormal ringdown for perturbations of the electromagnetic field.

Fig.\ref{fig6}-Fig.\ref{fig7} show that the scalar field perturbation and the electromagnetic perturbation of the charged black-bounce spacetimes for the case of $(\rvert {Q}\rvert \le m)$ and $(\rvert {l}\rvert > m+ \sqrt{m ^{2}-Q^{2} })$, these pictures present the following characteristics:

(1) In terms of the values of the bounce parameter $l$, the effective potentials (Fig.\ref{fig2} and Fig.\ref{fig3}) express two relatively distant peaks. This phenomenon causes modification of the signal at the late times labeled as echoes, which may generate either because of the modification of the black hole metric in the vicinity of its event horizon\cite{ret45,ret46}, or by reason of the matter situated at some distance from the compact object\cite{ret47,ret48}. After the black hole/wormhole transition, we find the quasinormal ringdown of the initial black hole followed by a series of echoes in Fig.\ref{fig6} and Fig.\ref{fig7}, and with the parameter $l$ increases, the time interval for the echo signals becomes shorter, which shows that the time interval has a strong dependence on the bounce parameter $l$, therefore we can see that when the bounce parameter $l$ reaches a certain level (Fig.\ref{fig6} and in Fig.\ref{fig7}), the echoes signals are not likely to be detected. Actually, these conclusions can be predicted from the effective potential in Fig.\ref{fig2}-Fig.\ref{fig3}, since as the parameter $l$ increases, the width of the potential well becomes smaller and smaller, and finally merges into a single peak potential barrier, thus the echoes signals ultimately vanish.

(2) When the bounce parameter $l$ become larger, the amplitudes of the echoes signal only slightly change, this result is probably due to the corresponding potential function. We can see that as the parameter $l$ increases, the peak value of the potential well only slightly change(such as Fig.\ref{fig2} and Fig.\ref{fig3}). It is because the potential well strongly constrain the scalar wave or electromagnetic wave that the echoes are generated. So almost the same height potential wells have the  almost the same probability for the scalar wave or electromagnetic wave escaping from the potential well, this is also the reason why the amplitudes of the echoes signal remain almost unchanged.

(3) As the bounce parameter $l$ increases, both peaks of the effective potential become closer and the echo effect decreases rather quickly, and the wormhole still remains a distinctive quasinormal mode. Moreover, from Fig.\ref{fig6}(c) and Fig.\ref{fig7}(c), we find that the time evolution of the traversable wormhole have a kind of the initial outburst and obvious quasinormal mode of the wormhole spacetime.

Fig.\ref{fig8} shows the time evolution of the two-way traversable wormhole's $(\rvert {Q}\rvert>m)$ quasinormal ringing for perturbation of the scalar field and perturbation of the electromagnetic field with different spacetime parameters, and obviously there are no echoes signals, which can be predicted from Fig.\ref{fig4}, since there is no potential well in the figure. Moreover, we can see that the two-way traversable wormhole's quasinormal ringing will decay slower with the increasing bounce parameter $l$.

Fig.\ref{fig9}-Fig.\ref{fig10} show that the effective potential($V_{2}$) and the semi-logarithmic plot of time evolution of electromagnetic perturbations for the charged black-bounce spacetimes with different spacetime parameters, analogy to the cases of the scalar field perturbation and electromagnetic field perturbation($V_{1}$), the echoes signal only appears when $(\rvert {Q}\rvert \le m)$ and $(\rvert {l}\rvert > m+ \sqrt{m ^{2}-Q^{2} })$.

Finally, Fig.\ref{fig11}-Fig.\ref{fig12} express the time-domain profiles for scalar field and electromagnetic field($V_{1}$ and $V_{2}$) in charged black-bounce spacetimes, which show that the echoes signal only appears when $(\rvert {Q}\rvert \le m)$ and $(\rvert {l}\rvert > m+ \sqrt{m ^{2}-Q^{2} })$, and this result is independent on the azimuthal quantum number $l_{0}$.

\section{The QNM frequencies of the charged black-bounce spacetimes}\label{sec5}

In this section, we will study the time domain profiles of the charged black-bounce spacetimes, as the introduction in Sec.\ref{sec3}, the Prony method is used for extracting the QNM frequencies. As stated bofore, the frequency of the QNMs is a complex number, i.e.  $\omega=\omega_{Re}+i{\omega}_{Im}$, with $\omega_{Re}$ is the real oscillation frequency and ${\omega}_{Im}$ is proportional to the decay rate of a given mode. We list the QNM frequencies of the charged black-bounce spacetimes in terms of the different bounce parameter $l$. It should be noted that the extraction of frequencies from the profiles obtained with the help of time domain integration strongly depends on the temporal range which is determined to be the quasinormal ringing, so in order to have a better comparison with the  frequencies under different parameters, we set the almost same temporal range. The table presents the following three characteristics:

(1)For the case of the regular black hole, as the bounce parameter $l$ increase, the damping rate is also slighter increasing, which implies the quasinormal ringdown has a faster decay rate, while the bounce parameter $l$ is not sensitive to the decay rate, this result verifies the properties of the Fig.\ref{fig5} and Fig.\ref{fig9}(a).

(2)For the case of the traversable wormhole, as the bounce parameter $l$ increase, the imaginary part of the QNM frequencies is decreasing.

(3)For the case of the two-way traversable wormhole, for the smaller bounce parameter $l$, the bigger ${\omega}_{Im}$ corresponds to the faster decay rate of the QNMs, this result can be predicted from Fig.\ref{fig8}. In addition, for the bigger bounce parameter $l$, $\omega_{Re}$ becomes smaller, which implies the real oscillation frequency of the QNMs becomes smaller.

\begin{table}[htbp]
\renewcommand\thetable{}
\caption{The fundamental quasinormal mode for the charged black-bounce spacetimes}
\scalebox{0.8}{
\begin{tabular}{lllll}
\hline
\multicolumn{2}{l}{}               & the scalar field($l_{0}=1$)                       & the electromagnetic field($l_{0}=1$, $V_{1}$)               & the electromagnetic field($l_{0}=2$, $V_{2}$)                \\ \hline
\multicolumn{5}{l}{the regular black hole}                                                                                                                     \\ \hline
\multirow{3}{*}{Q=0.03} & l=0.0002 & 0.730913-0.251657i                     & 0.618978-0.210527i                     & 0.933746-0.154364i                      \\
                        & l=0.0005 & 0.732307-0.252514i                     & 0.61962-0.21222i                       & 0.937353-0.156014i                      \\
                        & l=0.0008 & 0.732193-0.253059i                     & 0.619814-0.212311i                     & 0.937581-0.156082i                      \\ \hline
\multicolumn{5}{l}{the traversable wormhole}                                                                                                                   \\ \hline
\multirow{4}{*}{Q=0.03} & l=1.01   & 0.57998-0.138985i,echo                 & 0.518414-0.128371i,echo                      & 0.740378-0.116516i,echo                       \\
                        & l=1.03   & 0.581143-0.135181i,echo                & 0.520627-0.125885i,echo                      & 0.739762-0.113407i,echo                       \\
                        & l=1.2    & 0.642162-0.115745i                     & 0.472732-0.142619i                      & 0.732515-0.0945726i                      \\
                        & l=1.7    & 0.602112-0.0836229i                    & 0.539532-0.088682i                     & 0.72527-0.0661795i                      \\ \hline
\multicolumn{5}{l}{the two-way traversable wormhole}                                                                                                           \\ \hline
\multirow{3}{*}{Q=0.7}  & l=0.9    & 0.807521-0.356207i                     & 1.64928-0.294751i                      & 1.09557-0.140297i                       \\
                        & l=1.2    & 0.903813-0.185995i                     & 1.02976-0.234232i                      & 0.921498-0.109263i                      \\
                        & l=1.5    & 0.761068-0.162117i                     & 0.80138-0.183107i                      & 0.805344-0.0990703i                     \\ \hline
\end{tabular}  }
\end{table}

\section{Conclusions}\label{sec6}
In summary, the charged black-bounce spacetimes (Eq.\ref{eq1}) neatly interpolated between the regular black hole and traversable wormholes, we studied the ringing of the black hole/wormhole transition, which showed that this transition is characterized by echoes and observed three qualitatively distinct types of quasinormal ringdown behaviour:

 (i) the usual ringing of the regular black hole;

 (ii) the remnant of the initial fundamental quasinormal mode, followed by the series of echoes after the blackhole/wormhole transition;

 (iii) the obvious wormhole¡¯s mode, after the initial outburs.

Moreover, the QNMs of black holes are proper oscillations of black holes under different boundary conditions, which corrsponds to  purely incoming waves at the event horizon and purely outgoing waves at infinity, and the boundary conditions for a traversable wormhole  are the same in terms of the tortoise coordinate, therefore the alternative tools used for finding black holes QNMs can, in some sense,  be used for wormholes as well after some necessary corrections \cite {ret36,ret37,ret49,ret50,ret51}. This work we considered the scalar and electromagnetic perturbations for the charged black-bounce spacetimes, which showed that the echoes signals of the two perturbations have very similar characteristics. For future work, we can go further to investigate the gravitational field perturbations\cite {ret52} for the charged black-bounce spacetimes, and this study based on QNMs may be checked in future gravitational wave plans.

\bmhead{Acknowledgments}

The authors are grateful to Dr. Alexander Zhidenko from Brazil for providing us his mathematica code with the implementation of the Prony method. This research was funded by the Guizhou Provincial Science and Technology Project(Guizhou Scientific Foundation-ZK[2022] General 491) and National Natural Science Foundation of China (Grant No.11465006).

\bibliography{sn-bibliography}% common bib file

\end{document}